\documentclass[sigconf, authorversion,nonacm]{acmart}

\usepackage{tikz,graphics,color,fullpage,float,caption,subcaption}
\usepackage{multirow}
\graphicspath{{figures/}}
\usepackage{listings}

\copyrightyear{2023} 
\acmYear{2023} 
\setcopyright{rightsretained} 
\acmConference[SC-W 2023]{Workshops of The International Conference on High Performance Computing, Network, Storage, and Analysis}{November 12--17, 2023}{Denver, CO, USA}
\acmBooktitle{Workshops of The International Conference on High Performance Computing, Network, Storage, and Analysis (SC-W 2023), November 12--17, 2023, Denver, CO, USA}
\acmDOI{10.1145/3624062.3624239}
\acmISBN{979-8-4007-0785-8/23/11}

\begin{document}

\title{Analysis and Characterization of Performance Variability for OpenMP Runtime}

\author{Minyu Cui}
\orcid{1234-5678-9012}
\affiliation{%
  \institution{Chalmers University of Technology }
  \streetaddress{Gothenburg, Sweden}
  \city{Gothenburg}
  \country{Sweden}
  \postcode{43017-6221}}
  \email{minyu@chalmers.se}

\author{Nikela Papadopoulou}
\affiliation{%
  \institution{Chalmers University of Technology}
  \city{Gothenburg}
  \country{Sweden}}
\email{nikela@chalmers.se}

\author{Miquel Peric\`as}
\affiliation{%
  \institution{Chalmers University of Technology}
  \city{Gothenburg}
  \country{Sweden}}
\email{miquelp@chalmers.se}

\renewcommand{\shortauthors}{M. Cui et al.}

\begin{abstract}
In the high performance computing (HPC) domain, performance variability is a major scalability issue for parallel computing applications with heavy synchronization and communication.
In this paper, we present an experimental performance analysis of OpenMP benchmarks regarding the variation of execution time, and determine the potential factors causing performance variability.
Our work offers some understanding of performance distributions and directions for future work on how to mitigate variability for OpenMP-based applications. Two representative OpenMP benchmarks from the EPCC OpenMP micro-benchmark suite and BabelStream are run across two x86 multicore platforms featuring up to 256 threads. From the obtained results, we characterize and explain the execution time variability as a function of thread-pinning, simultaneous multithreading (SMT) and core frequency variation. \looseness=-1

\end{abstract}

\keywords{performance variability, OpenMP, parallel computing, thread-pinning, simultaneous multithreading}


\maketitle
\pagestyle{plain} 

\section{Introduction}
\label{Intro}
%
Parallel applications executing on shared-memory systems in the HPC world usually follow the single-program multiple-data (SPMD) model, typically implemented with OpenMP. OpenMP, the de facto programming model for SPMD, spawns multiple threads when encountering a \texttt{\#pragma omp parallel} clause. 
Each thread is then executed on one core/hardware thread of the system to execute the parallel region, and commonly all threads synchronize at the end of the execution of the parallel region to compute the final result.
Some system-specific activities, such as operating system (OS) daemons and interrupt processing, can cause preemption or interrupt handling to one or multiple of the threads, causing the execution of the parallel work to be delayed and the execution time to be dominantly determined by the slowest thread, while the others wait for synchronization, leading to a waste of resources like time and energy. 
Also, due to the randomness of the delay, it will in turn generate performance variability for the runtime of the parallel application. 
Performance variability has become an important limiter to the scalability in parallel computing~\cite{Kocoloski2018}. With the complexity of modern hardware architecture features increasing, variability has become an increasingly challenging issue for improving the efficiency of parallel computing~\cite{Marathe2017}. 

Performance variability or run-to-run variations of applications owing to multiple components in the system can become an obstacle for the development of parallel applications in several ways, like performance debugging or quantifying the effects of system software and compilers changes~\cite{Bhatele2013ThereGT}. There have been various efforts to identify the potential causes of variability and in turn find solutions to reduce the possibility of variability occurrence, aiming at obtaining performance stability of parallel application executions. \looseness=-1
Most studies on variability have focused on MPI~\cite{Gerofi2021,Inadomi2015,Leon2016,Morari2011}, as the explicit and synchronizing nature of message passing communication, and the large scale of applications using this programming model, make MPI applications more sensitive to noise, which in turn leads to load imbalance and ultimately performance degradation. Evaluating the impact of noise and the occurring performance variability in shared memory models such as OpenMP has received comparatively less attention. However, as the core count of modern CPUs increases, shared memory parallel applications using OpenMP are likely to be also impacted by OS noise.
 

Several strategies to optimize the performance of parallel programs with OpenMP have been proposed, studied, and have influenced the state-of-practice in execution. A common strategy is thread pinning~\cite{HPCWiki2022}, which can improve application performance by keeping threads bound to a specific core and avoiding expensive memory accesses.  The authors in~\cite{Mazouz2011} have studied several thread-pinning strategies to improve the performance of OpenMP programs, while in a later work~\cite{Mazouz2013}, they have proposed dynamic thread-pinning for phase-based OpenMP programs with multiple parallel regions. Their study is limited to a small scale of core/thread counts. However, they identified thread pinning as a critical factor to performance variability~\cite{mazouz2011analysing}.
The effective usage of simultaneous multithreading, the architectural mechanism that supports several hardware threads per physical core, has also been shown to improve the performance of MPI and MPI+OpenMP applications~\cite{Leon2016}.
Finally, tuning the core frequencies for performance is also important, and well-studied in the literature relevant to developing dynamic energy-efficiency techniques~\cite{chen2022steer,li2012strategies, marathe2015run}, as even in steady-state, frequency variation can cause high variability of the performance~\cite{porterfield2013openmp}. 

This work focuses on characterizing the performance variability of OpenMP on modern CPUs. Motivated by the scale of recent, modern multi-core systems, we conduct an extensive study of the impact of common performance-optimizing strategies on the performance of OpenMP applications, in an effort to further understand and pinpoint sources of performance variability in OpenMP. We use two micro-benchmarks from the EPCC benchmark suite~\cite{lagrone2011set} which focus on the performance of common OpenMP constructs, such as \texttt{parallel for} and synchronization, and the BabelStream benchmark~\cite{deakin2018evaluating}, which assesses the memory bandwidth, and execute them on two different systems, using different variability-reducing strategies. In particular, 
we analyze thread-pinning, which can help in revealing the performance degradation related to unbound threads in parallel applications. We additionally explore simultaneous multithreading to show how it can affect the performance of OpenMP benchmark executions. Finally, during benchmark execution, we record the frequencies of all cores, to examine whether frequency variation exists, and how it can affect the variability of execution time.  

Our study is conducted on two production clusters hosted by two different academic institutions. We do not have privileged access to these systems and therefore cannot control the node setup or operating system knobs and we cannot trace kernel-level events. 
We, therefore, rely on a statistical analysis of the observed execution times, repeatedly running every benchmark with multiple iterations of the kernels of interest. By studying the possible sources and characterizing the performance variability, we 
can categorize the sources of performance variability and find efficient solutions to mitigate a particular class of variability in future work.  

The rest of the paper is organized as follows. We introduce related work in Section~\ref{Related_work}. Section~\ref{Methodology} provides an overview of the proposed methodology to characterize the performance variability. We present our experimental setup in Section~\ref{setup} and experimental results in Section~\ref{EX_results}. A conclusion of this work follows in Section~\ref{conclusion}.


\section{Related work}
\label{Related_work}
Performance variability of parallel applications has been well reported on modern systems in multiple works. 
At the extra-application level, there can be multiple reasons for unpredictable performance, with operating system noise (also referred to as \textit{OS jitter}) being one of the most common reasons. Several works~\cite{Gerofi2021,Morari2011,de_Oliveira2023,Tsafrir2005} study the impact of operating system activities on performance, looking primarily at large-scale parallel applications with MPI. A recent work~\cite{Song2021} demonstrates that OS noise on non-uniform memory access (NUMA) architectures can cause high run-to-run performance variability. 
As the number of cores/processors on modern systems grows, OS noise can become a more significant factor of performance variability, as a small amount of perturbation can be greatly amplified in parallel computing. It is therefore important to study the impact of OS noise on performance variability and find solutions to mitigate it.
%
%

Aside from the operating system, performance variability can arise from contention and interference on shared resources. Bhatele et al.~\cite{Bhatele2013ThereGT} show that sharing network resources on HPC systems is a primary source of performance variability. Xu et al.~\cite{LiXu2021} show that interference on the I/O subsystem affects the performance of parallel applications. On systems with simultaneous multithreading, performance degradation can occur from oversubscription of the physical cores~\cite{Leon2016}. Another source of variability is manufacturing variability~\cite{Inadomi2015}, which leads to performance heterogeneity. The power variation from manufacturing variability can affect the performance stability of HPC applications, as it translates to CPU frequency variation~\cite{Rountree2012}. 


As there is increasing evidence for performance variability of parallel applications, several techniques and tools have been proposed to measure and characterize performance variability in recent works. 
In particular, for OS noise, Pradipta et al.~\cite{Pradipta2007} develop a tool to monitor and evaluate the impact of OS noise on Linux-based systems through fine-grained kernel instrumentation. Gioiosa et al.~\cite{Gioiosa2004} extend \textit{Oprofile}, a Linux kernel-level tool to characterize the sources of OS noise. Morari et al.~\cite{Morari2011} extend the Linux tool \textit{LTTng} to build \textit{LTTng-Noise}, a tracing tool. De Oliveira et al.~\cite{de_Oliveira2023} develop the \textit{osnoise} tracer, which analyzes noise activities via kernel instrumentation.
A more generic technique to measure performance variability and statistically characterize performance distributions has been proposed by Kocoloski et al.~\cite{Kocoloski2018}, to assist in system parameter design such as power-capping. 
%

A limited number of works have focused on analyzing the performance variability of OpenMP programs. 
Camacho et al.~\cite{Camacho2016} show that thread binding can reduce execution time variation in OpenMP applications, and Mazouz et al.~\cite{mazouz2011analysing} study the effects of thread binding, OS jitter, and hardware-related sources (memory-access related sources, concurrent jobs, asymmetry between cores, dynamic voltage scaling and device temperature) on execution time variation in OpenMP.
In our work, we also focus on analyzing and characterizing performance variability in OpenMP. We exclude interference from other applications and run our benchmarks in isolation. Additionally, as we do not have privileged access to the platforms in study, we exclude operating system knobs from our techniques and only observe the impact of operating system noise on OpenMP, with a statistical analysis of results. 


\section{Methodology}
\label{Methodology}
In this section, we describe the methodology followed to characterize performance variability in OpenMP. We note that we always execute benchmarks in isolation on a single node, eliminating the case of variability from application interference. We perform our experiments on production, site-managed clusters, therefore we do not have privileged access that would allow us to tune the execution environment. Instead of detailed trace analysis, we rely on multiple experiments and statistical analysis of the results. The following paragraphs describe the strategies we apply to detect the sources and impact of performance variability.\looseness=-1 

\textbf{Thread pinning:}
By default, we let the operating system decide the thread placement on cores, as the default setup for \texttt{OMP\_PROC\_BIND} is set to false. In this case (before thread-pinning), the threads may migrate between cores during the execution of parallel programs to improve work balance. 
In modern multi-core architectures, exploitation of locality is essential to efficiently run parallel programs~\cite{Iwasaki2019}.
OpenMP supports users with fine-grained thread affinity control through thread pinning. A group of \textit{places}, corresponding to a group of hardware threads, can be defined and OpenMP threads can be bound to specific places (therefore hardware threads) by setting a pinning policy. To achieve this, some OpenMP-related environment variables are used to specify the OpenMP settings in this paper, i.e., \texttt{OMP\_NUM\_THREADS} is used to define the number of threads, \texttt{OMP\_PLACES} and \texttt{OMP\_PROC\_BIND} work together to pin each thread to a specific core. The thread affinity policy is set as \textit{close}, which implies that worker threads are close to the main thread in contiguous partitions~\cite{HPCWiki2022,OpenMP-API2018}. 

\textbf{Using Simultaneous Multithreading:}
Simultaneous multithreading mechanism is implemented on one of the two platforms included in our experimental setup, Dardel (see Section~\ref{setup} for details), where each core has two hardware threads (also referred to as logical cores). We evaluate two configurations in our experiments. The first one is single-threaded, denoted as \textbf{ST}, in which at most one hardware thread per physical core is used to run the benchmark. In this case, the additional hardware thread of the core is reserved for operating system activities to absorb noise and isolate the benchmark running from the system interference. In the second configuration, both hardware threads of the core are utilized to run our benchmarks. We refer to this configuration as \textbf{MT}.
We collect the results under the above two configurations and compare the performance variability of the execution of the benchmarks. \looseness=-1

\textbf{Frequency logging on a separate core:} 
During the execution of benchmarks, a background Python script is run on a separate core to collect the frequencies of all cores. By doing this, we try to avoid interference from the frequency logger and benchmark running on the same core and guarantee that the execution of benchmarks is influenced as little
as possible by other background activities. In the next section, we showcase the performance variability that can be related to the frequency variations. \looseness=-1

\section{Experimental Setup}
\label{setup}
\subsection{Hardware platforms}

We use two different hardware platforms for our experiments. The first platform, \textit{Dardel}, is an HPE Cray EX supercomputer located at the PDC Center for High-Performance Computing in Sweden. Each node of Dardel integrates two AMD EPYC Zen2 2.25GHz 64-core processors, accommodating two hardware threads per core. From the operating system's view, there are a total of 128 cores and 256 hardware threads/logical cores. The cores are organized in 8 NUMA domains of 16 cores each, with each socket behaving as a quad-NUMA domain. The maximum frequency of each core is 3.4GHz. The system runs the SUSE Linux Enterprise Server 15 SP3 OS, with Linux kernel version 5.3.18-150300.59.76\_11.0.53-cray\_shasta\_c. We use \texttt{gcc v7.5.0} as the compiler.  

The second platform, \textit{Vera}, is a cluster located at C3SE Center for Scientific and Technical Computing at Chalmers University of Technology in Sweden. Each node of Vera integrates two Intel Xeon Gold 6130 2.1GHz 16-core processors, with a total of 32 cores. Each socket corresponds to a NUMA domain, with a total of 2 NUMA domains on the node. The maximum frequency of each core is 3.7GHz. The system runs Rocky Linux release 8.7, with Linux kernel version 4.18.0. We use \texttt{gcc v8.5.0} as the compiler.

\subsection{OpenMP benchmarks}
We use three different OpenMP benchmarks for our evaluation of performance variability in OpenMP. We draw two benchmarks from the EPCC OpenMP micro-benchmark suite \cite{LaGrone2011ASO, WangP2021}, one of the most comprehensive suites for OpenMP constructs, which provides measurements of the overhead incurred from an OpenMP construct by comparing the execution time of parallel code against this of serial code. We use \textit{schedbench}, the benchmark focusing on the \texttt{parallel for} construct with different schedules, and \textit{syncbench}, the benchmark which evaluates all the different available synchronization methods in OpenMP. The benchmarks can be run with different parameters. We present the parameters used for the two benchmarks in our evaluation in Table~\ref{tbl:benchmark_EPCC}.
%
The third benchmark is \textit{BabelStream} \cite{Deakin2018}, a common benchmark to measure memory bandwidth by executing simple vector kernels, including copy, add, multiplication, triad, and dot product. It has been used in previous work~\cite{Inadomi2015} to evaluate performance variability in a power-limited environment. We use the default parameters and an array size of $2^{25}$ for \textit{BabelStream} in our evaluation. 
 

\begin{table}[]
\caption{Parameters of the EPCC OpenMP micro-benchmarks}
\vspace{-8pt}
\label{tbl:benchmark_EPCC}
\begin{tabular}{ccc}
 \hline
EPCC micro-benchmark & schedbench & syncbench  \\   
\hline
{outer repetitions} & 100 &100 \\ 
{delay time(\textmu s)}& 15 &0.1  \\ 
{test time(\textmu s)} & 1000 &1000\\ 
{itersperthr} & 8192 & - \\ 
\hline
\vspace{-12pt}
\end{tabular}

\end{table}


We have executed a large set of experiments with the three benchmarks on the two hardware platforms described above. For every runtime configuration, we run each experiment 10 times, to collect run-to-run performance variability, in addition to any variability reported by the EPCC benchmarks themselves, which also execute 100 repetitions of each micro-benchmark. Due to page limitations, we only highlight those experimental results that show statistically significant performance variability and can shed light on the potential sources of this variability. 
In particular, we execute \textit{schedbench} with three different schedules, namely \texttt{static}, \texttt{dynamic} and \texttt{guided} and various different chunk sizes~\cite{Bull2007}, and present the results for specific schedules with the chunk size equal to \texttt{1}. e.g., \texttt{static} or \texttt{dynamic} schedule with chunk size equal to \texttt{1}, labelled as \texttt{static\_1} and \texttt{dynamic\_1} respectively.
From \textit{syncbench}, we select the \texttt{reduction} clause as the most representative of synchronization methods in OpenMP.  
For \textit{BabelStream}, in every single run, we collect the minimum, average, and maximum execution time for each kernel and then normalize the minimum and maximum execution time to the average execution time. The run-to-run variations of execution time are depicted by comparing the normalized minimum and maximum execution times among 10 runs for every vector operation kernel respectively.\looseness=-1

\section{Experimental Results}
\label{EX_results}
\subsection{OpenMP scalability}
\begin{table}[]
\caption{Higher execution time (\textmu s) for \textit{schedbench} (\texttt{dynamic\_1}). }
\label{tbl:higher_exetime_increasing_cores_sched_dynamic1}
\scalebox{0.9}{ 
\vspace{-8pt}
\begin{tabular}{ccccc}
\hline
\multirow{2}{*}{run \#} & \multicolumn{2}{c}{Dardel}                               & \multicolumn{2}{c}{Vera}                                \\ \cline{2-5}   & \multicolumn{1}{c}{4 threads}         & \multicolumn{1}{c}{254 threads} & \multicolumn{1}{c}{4 threads}         & \multicolumn{1}{c}{30 threads} \\ \hline
1                                                                          & \multicolumn{1}{l}{124020.18} & 154277.48                & \multicolumn{1}{l}{136485.84} & 164672.22               \\ 
2                                                                          & \multicolumn{1}{l}{124062.15} & 154162.74                & \multicolumn{1}{l}{136562.99} & 164642.42               \\ 
3                                                                          & \multicolumn{1}{l}{123989.57} & 154159.44                & \multicolumn{1}{l}{136621.25} & 164662.94               \\ 
4                                                                          & \multicolumn{1}{l}{123949.12} & 153999.99                & \multicolumn{1}{l}{136509.92} & 164665.88               \\ 
5                                                                          & \multicolumn{1}{l}{124016.11} & 154206.72                & \multicolumn{1}{l}{136386.38} & 164652.30               \\ 
6                                                                          & \multicolumn{1}{l}{123917.89} & 154044.95                & \multicolumn{1}{l}{136479.29} & 164573.27               \\ 
7                                                                          & \multicolumn{1}{l}{123885.41} & 154222.61                & \multicolumn{1}{l}{136513.89} & 164699.42               \\ 
8                                                                          & \multicolumn{1}{l}{123902.87} & 154182.23                & \multicolumn{1}{l}{136448.17} & 164754.22               \\ 
9                                                                          & \multicolumn{1}{l}{123935.31} & 168835.06                & \multicolumn{1}{l}{136645.71} & 164717.21               \\
10                                                                         & \multicolumn{1}{l}{124023.24} & 154065.79                & \multicolumn{1}{l}{136743.01} & 164757.37               \\ \hline
\end{tabular}
}
\vspace{-10pt}
\end{table}

\begin{figure}[htbp]
\vspace{-8pt}
\centering
\begin{subfigure}[b]{0.4\textwidth}
{\includegraphics[width=1.0\textwidth,trim=8 8 5 10, clip]{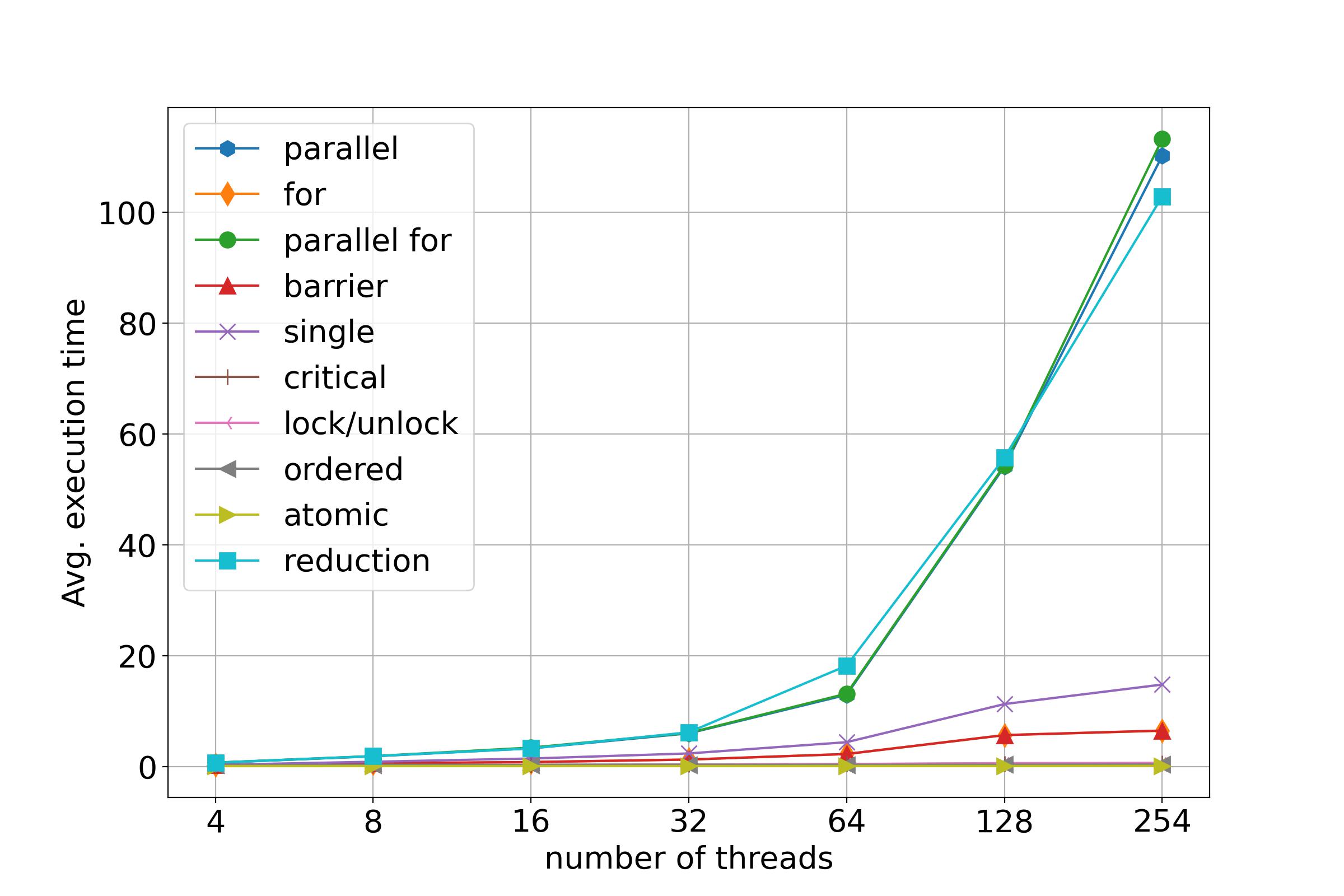}
\caption{4-254 threads on Dardel}
\label{fig:dardel_sync_exetime_all_core}}
\end{subfigure}
\begin{subfigure}[b]{0.4\textwidth}
{\includegraphics[width=1.0\linewidth]{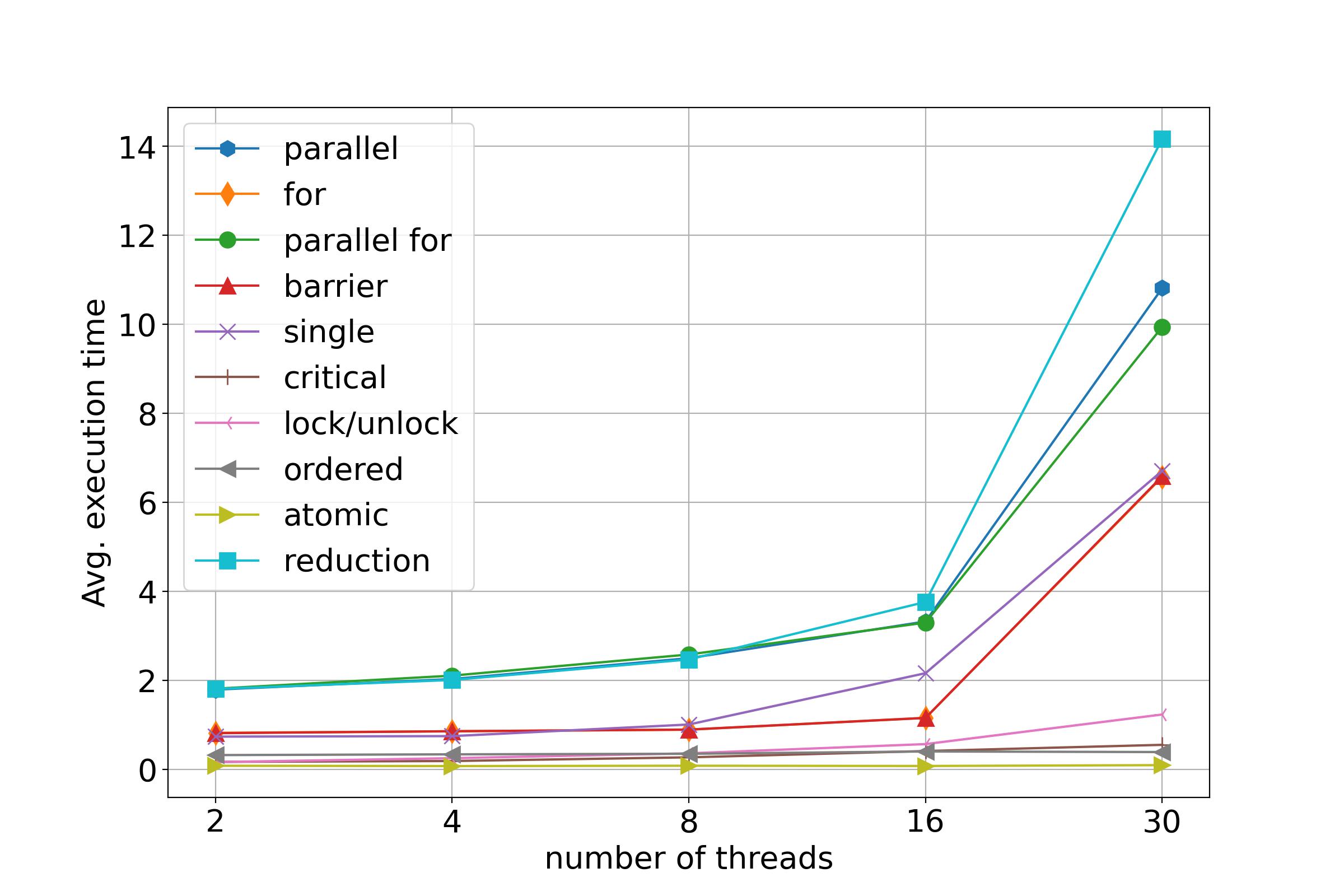}
\caption{2-30 threads on Vera}
\label{fig:vera_sync_exetime_all_core}}
\end{subfigure}
\caption{Execution time (\textmu s) when increasing the number of HW threads in \textit{syncbench} on Dardel and Vera.}
\label{fig:higher_exetime_increasing_cores_sync}
\vspace{-10pt}
\end{figure}

\begin{figure}[htbp]
\centering
\begin{subfigure}[b]{0.4\textwidth}
{\includegraphics[width=\textwidth]{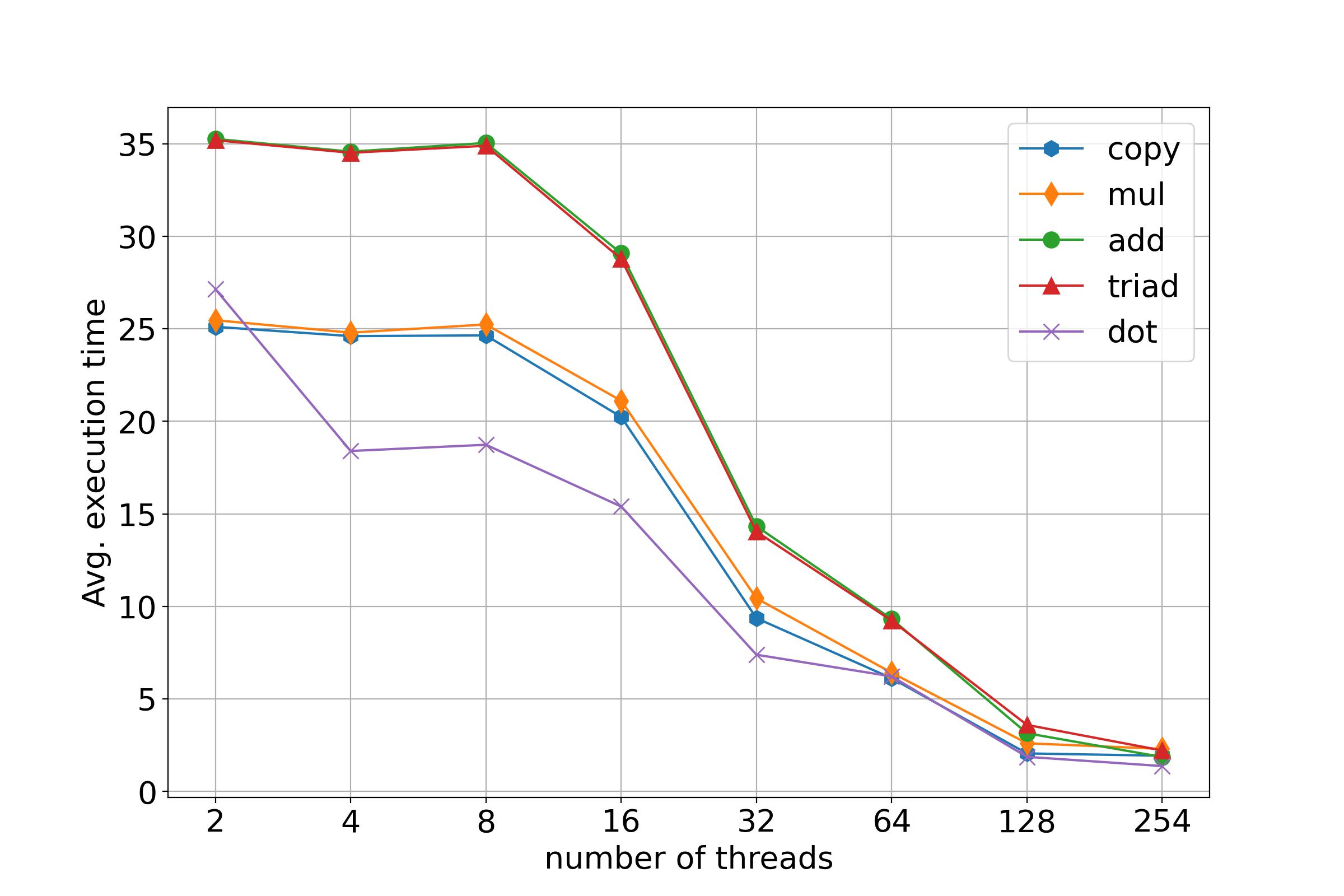}
\caption{2-254 threads on Dardel}
\label{fig:dardel_bstream_exetime_all_core}}
\end{subfigure}
\begin{subfigure}[b]{0.4\textwidth}
{\includegraphics[width=\textwidth]{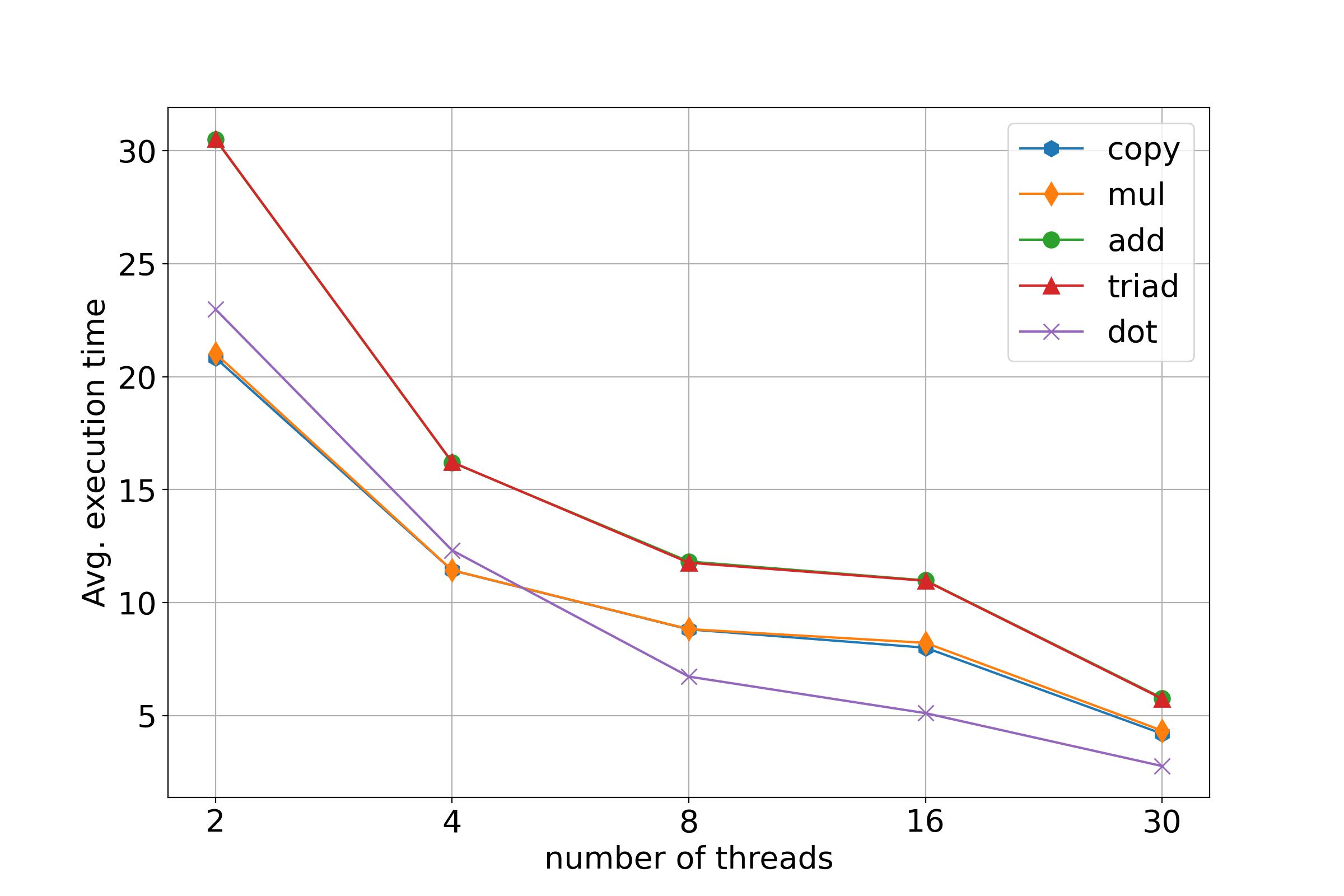}
\caption{2-30 threads on Vera}
\label{fig:vera_bstream_exetime_all_core}}
\end{subfigure}
\caption{Execution time (ms) for \textit{BabelStream} when increasing the number of HW threads on Dardel and Vera.}
\label{fig:lower_exetime_increasing_cores_BStream}
\vspace{-10pt}
\end{figure}

\begin{figure*}[ht]
\centering

\begin{subfigure}[b]{0.3\textwidth}
{\includegraphics[width=1.1\textwidth]{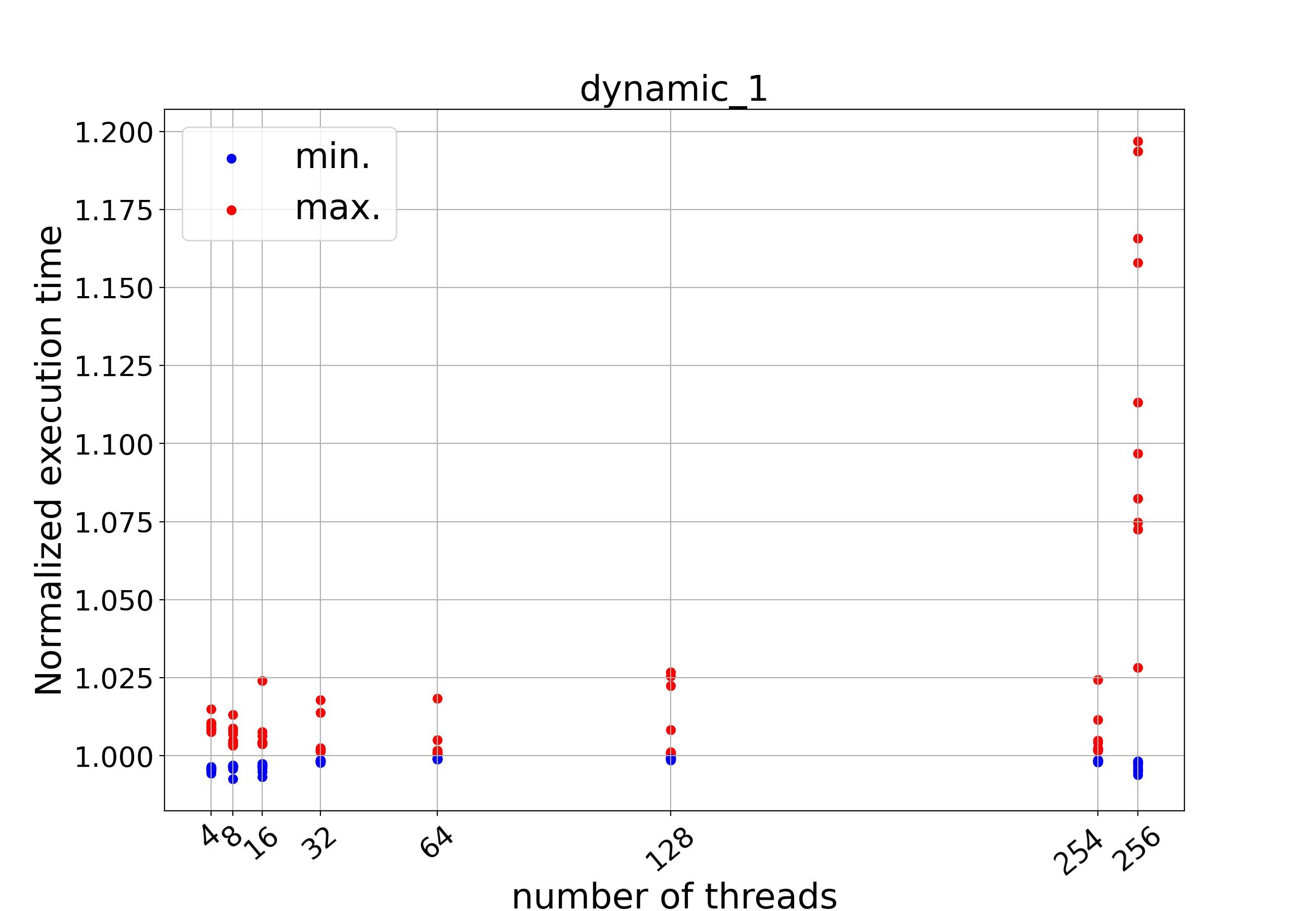}
\caption{\textit{schedbench} on Dardel }
\label{fig:dardel_dynamic1_normrized_min_max_exetime_scalability_10run_mt_allcore_256_dotplot}}
\end{subfigure}
\begin{subfigure}[b]{0.3\textwidth}
{\includegraphics[width=1.1\textwidth]{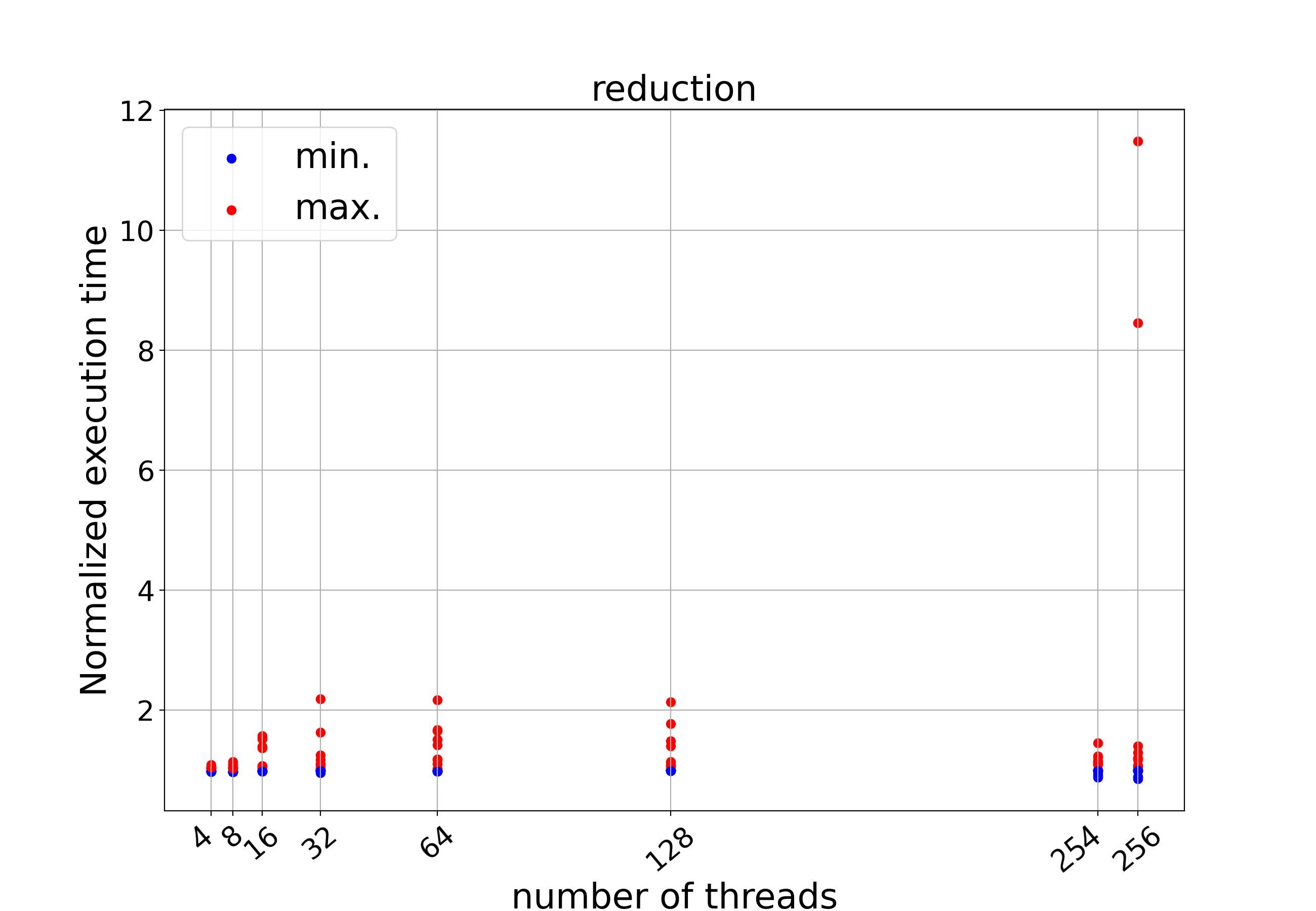}
\caption{\textit{syncbench} on Dardel }
\label{fig:dardel_reduction_normrized_min_max_exetime_scalability_10run_mt_allcore_256_dotplot_proportional}}
\end{subfigure}
\begin{subfigure}[b]{0.3\textwidth}
{\includegraphics[width=1.1\textwidth]{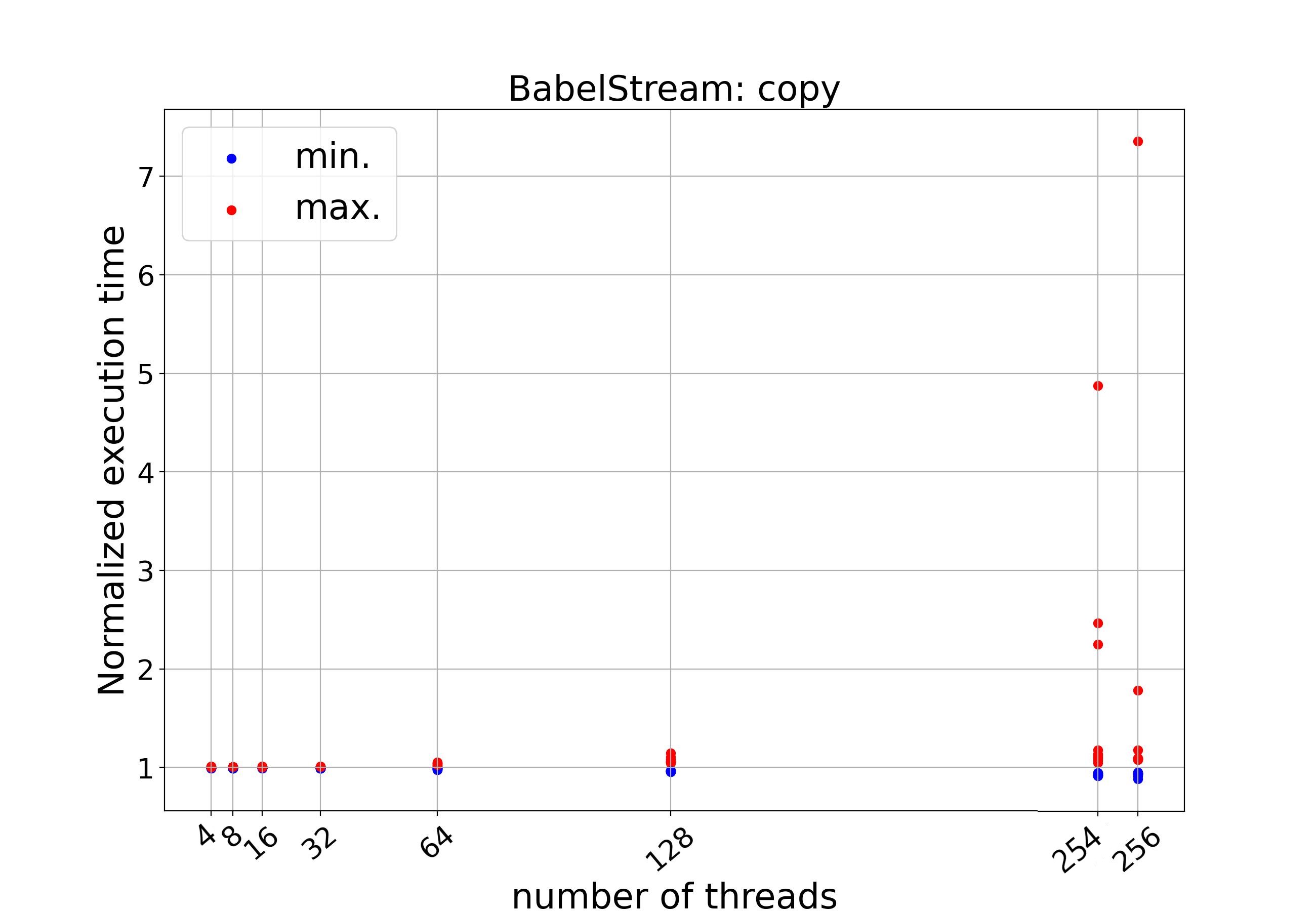}
\caption{\textit{BabelStream} on Dardel }
\label{fig:dardel_BStream_copy_normarized_min_max_scalability_10run_allcore_256_proportional}}
\end{subfigure}
\quad
\begin{subfigure}[b]{0.3\textwidth}
\centering
{\includegraphics[width=1.1\textwidth]{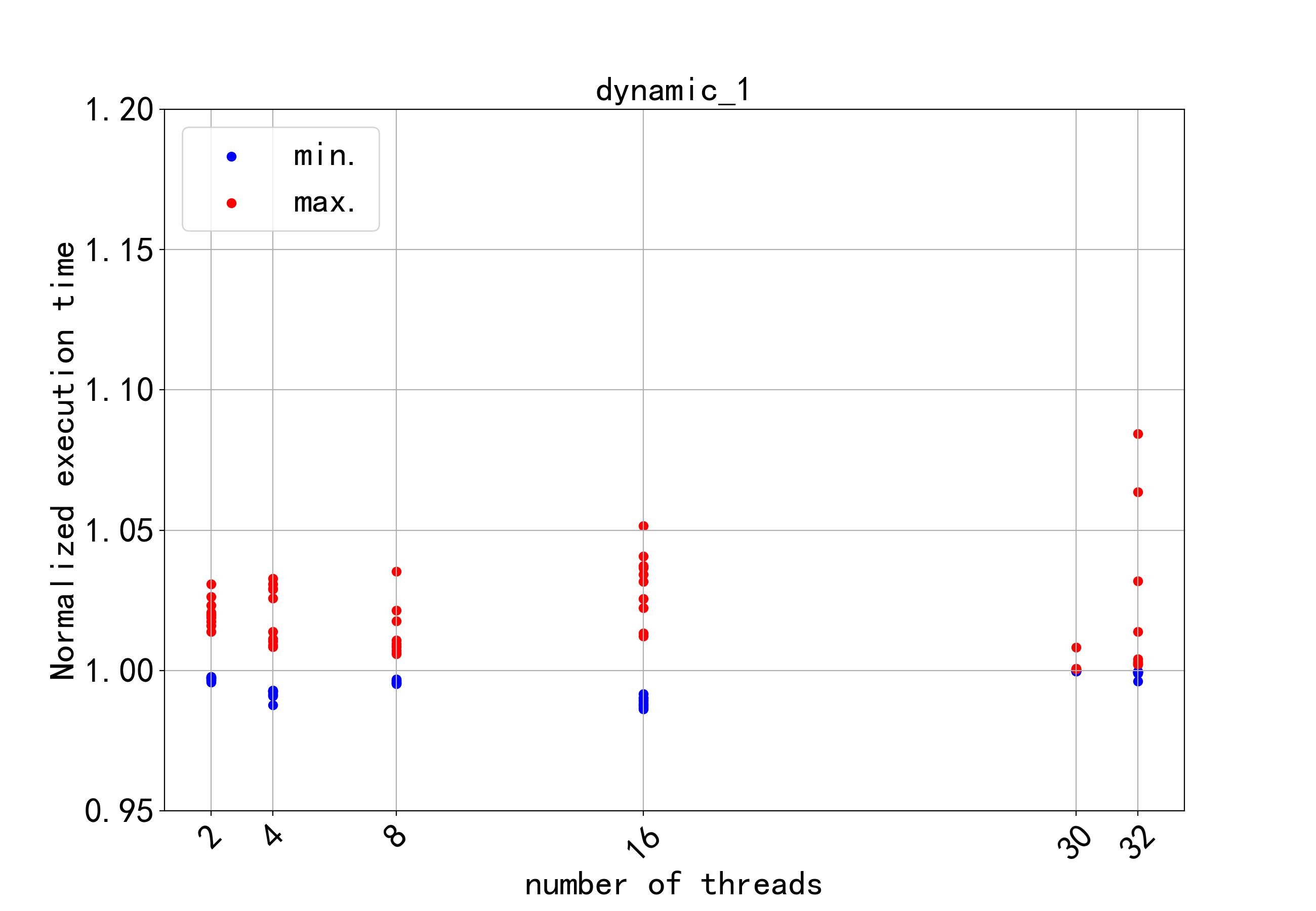}
\caption{\textit{schedbench} on Vera }
\label{fig:vera_dynamic1_normrized_min_max_exetime_scalability_10run_st_allcore_32_dotplot_proportional}}
\end{subfigure}
\begin{subfigure}[b]{0.3\textwidth}
\centering
{\includegraphics[width=1.1\textwidth]{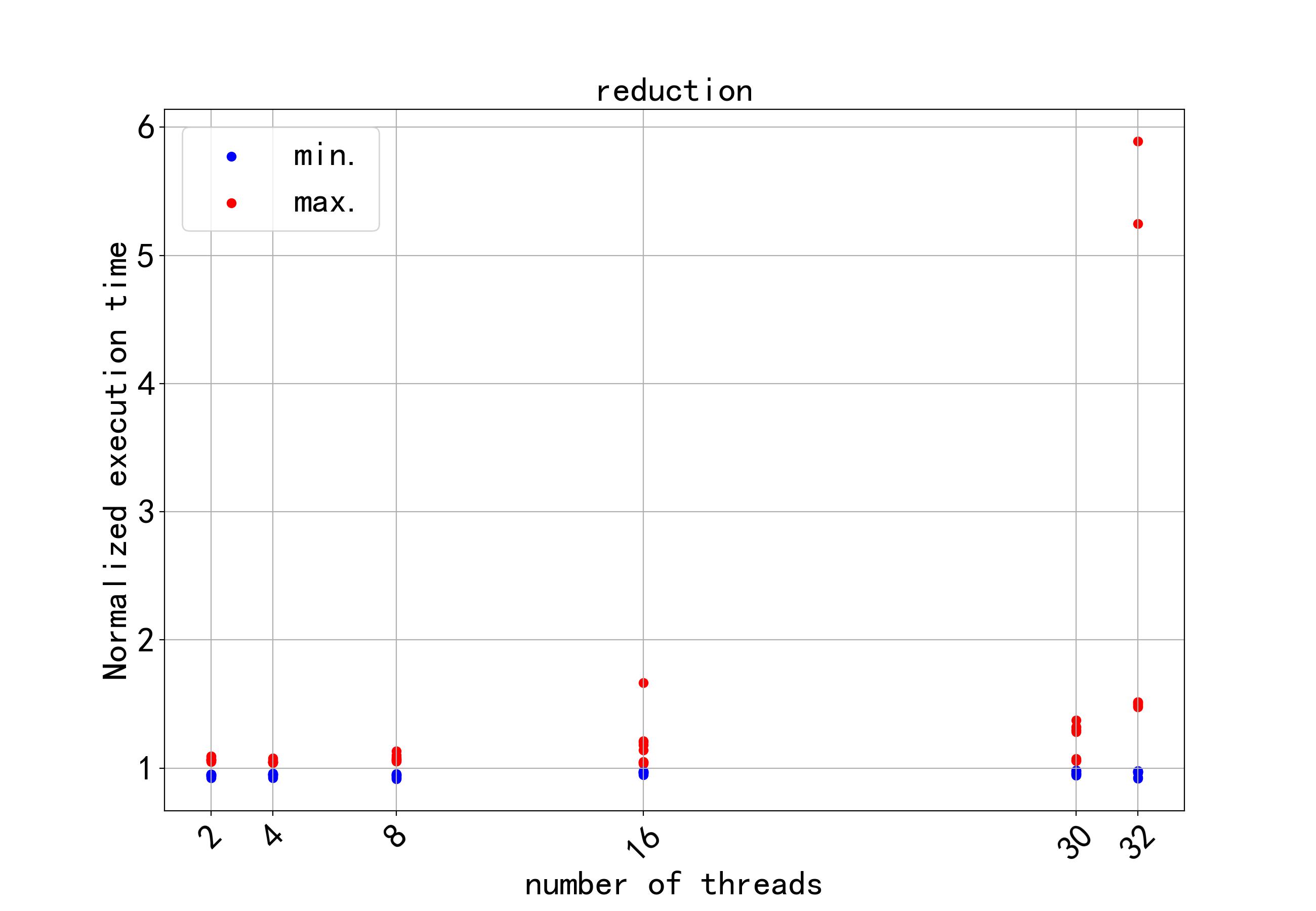}
\caption{\textit{syncbench} on Vera }
\label{fig:vera_reduction_normrized_min_max_exetime_scalability_10run_st_allcore_32_dotplot_proportional}}
\end{subfigure}
\begin{subfigure}[b]{0.3\textwidth}
{\includegraphics[width=1.1\textwidth]{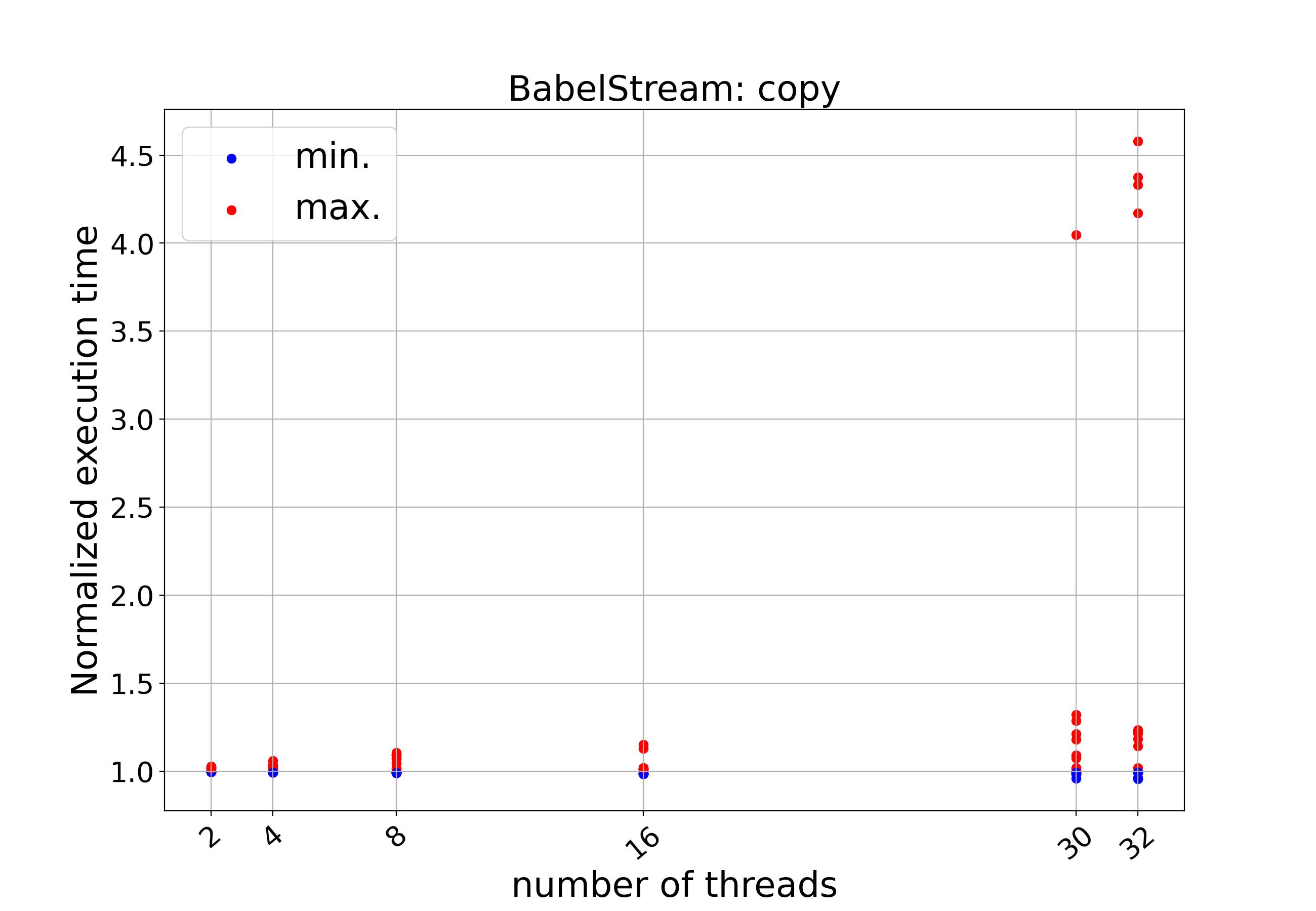}
\caption{\textit{BabelStream} on Vera }
\label{fig:vera_BStream_copy_normarized_min_max_scalability_10run_allcore_32_dotplot_proportional}}
\end{subfigure}
\caption{Scalability of performance variability of normalized execution time in \textit{schedbench}, \textit{syncbench}, and \textit{BabelStream} when increasing the number of used HW threads on Dardel and Vera.}
\label{fig:scalability_performance _variability_dotplot_proportional}
\vspace{-5pt}
\end{figure*}

We begin our evaluation by examining the scalability of the three OpenMP benchmarks, to understand the trend of the average (Avg.) execution time in Table~\ref{tbl:higher_exetime_increasing_cores_sched_dynamic1} and Figures~\ref{fig:higher_exetime_increasing_cores_sync} and~\ref{fig:lower_exetime_increasing_cores_BStream}, and examine whether higher thread counts have higher performance variability in Figure~\ref{fig:scalability_performance _variability_dotplot_proportional}. In Figure~\ref{fig:scalability_performance _variability_dotplot_proportional}, the minimum and maximum execution times are normalized to the average execution time for each run respectively, and run each benchmark 10 times.
We employ thread pinning for all the experiments, and make use of SMT, where available. \looseness=-1

Regarding the scalability of execution time, 
we observe that the execution time increases as we spawn additional OpenMP threads, for \textit{schedbench} in Table~\ref{tbl:higher_exetime_increasing_cores_sched_dynamic1} and for \textit{syncbench} in Figure~\ref{fig:higher_exetime_increasing_cores_sync}, showing the average execution time for all 10 runs.
For \textit{syncbench}, we additionally observe a sharp increase in the execution time when we start utilizing the second socket on both systems (30 threads with 2 NUMA domains for Vera, and on 128 threads with 2 quad-NUMA domains for Dardel), as well as when we utilize the logical cores, in addition to the physical cores, on Dardel (254 threads). Also, we additionally highlight that the micro-benchmark corresponding to the \texttt{reduction} clause is the most time-consuming among the synchronization micro-benchmarks.
We finally showcase the scalability of \textit{BabelStream} in Figure~\ref{fig:lower_exetime_increasing_cores_BStream}, observing that the execution time of \textit{BabelStream} reduces when launching more parallel threads, as expected, on both Dardel and Vera. 
\looseness=-1

Regarding the scalability of performance variability,   
higher thread counts add to performance variability for \textit{syncbench} and \textit{BabelStream} in Figure~\ref{fig:scalability_performance _variability_dotplot_proportional}, especially when the thread count is high ($\geq 128$ HW threads/OpenMP threads on Dardel and $\geq 30$ on Vera), while it is not as pronounced for \textit{schedbench},
as seen in the first column of Figure~\ref{fig:scalability_performance _variability_dotplot_proportional}.
%
%
It is worth pointing out that when all cores/HW threads were used for this scalability experiment, we observed a significantly worse performance behavior. To avoid this, on both systems, we spare 2 cores/HW threads, using 30 out of the 32 cores on Vera and 254 out of the 256 hardware threads on Dardel. 
We highlight that we observe both higher run-to-run variability, and also high variability between the 100 repetitions of the micro-benchmark for \textit{schedbench} and \textit{syncbench}. 
We argue that this is due to operating system activities, which interfere with the benchmark execution when no spare cores/resources are left for them, causing more noise from the view of the user's application execution. This observation is in accordance with recent works~\cite{Gerofi2021,de_Oliveira2023}, which argue for resource isolation required for system activities, in order to reduce the OS noise. \looseness=-1

\subsection{The effect of thread pinning}

\begin{figure*}[ht]
\vspace{-8pt}
\centering
\begin{subfigure}[b]{0.3\linewidth}
{\includegraphics[width=1.0\textwidth]{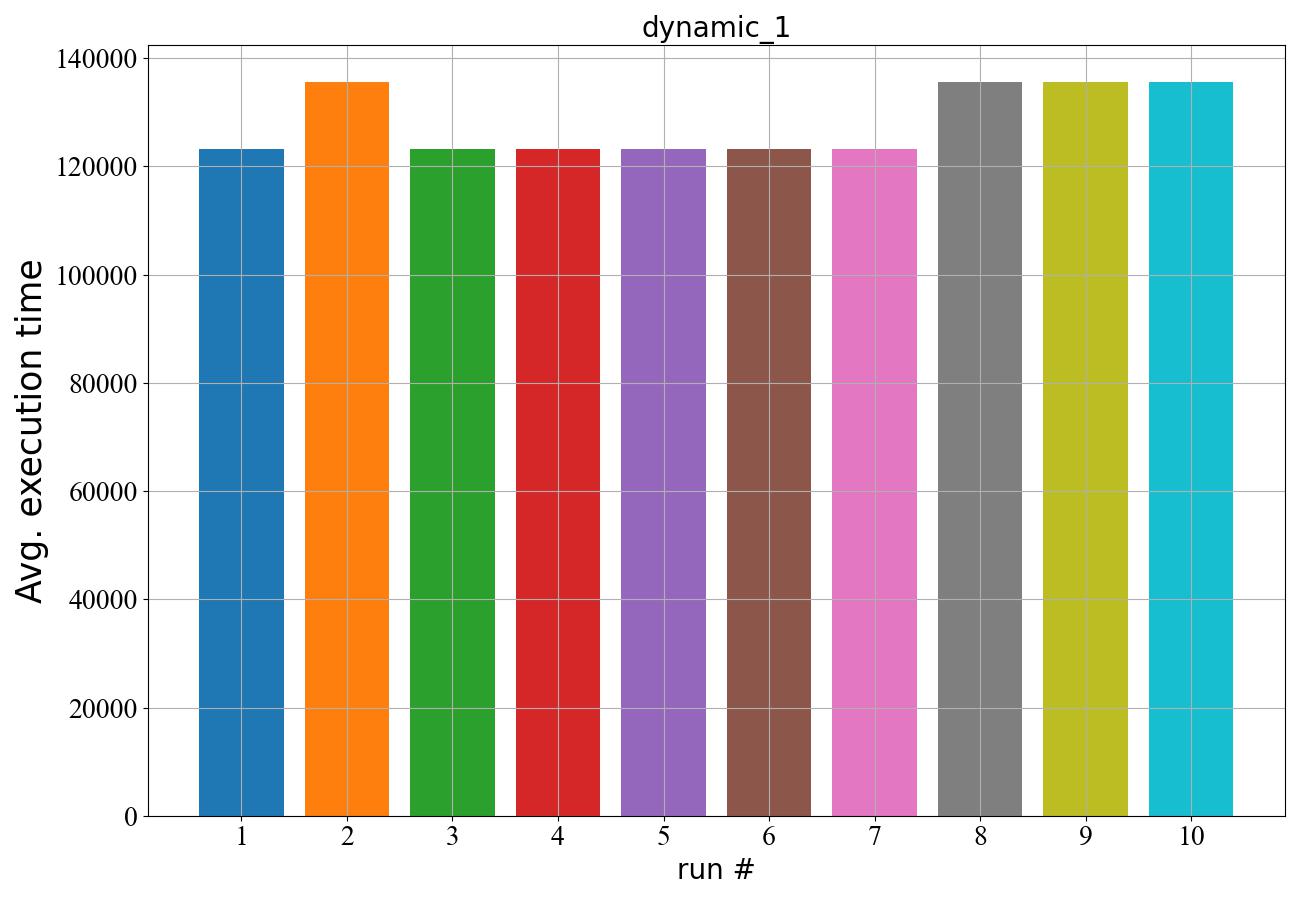}
\caption{\textit{schedbench} before thread-pinning}
\label{fig:dardel_dynamic_1_wopin_exetime_16core_st}}
\end{subfigure}
\begin{subfigure}[b]{0.3\linewidth}
{\includegraphics[width=1.1\textwidth]{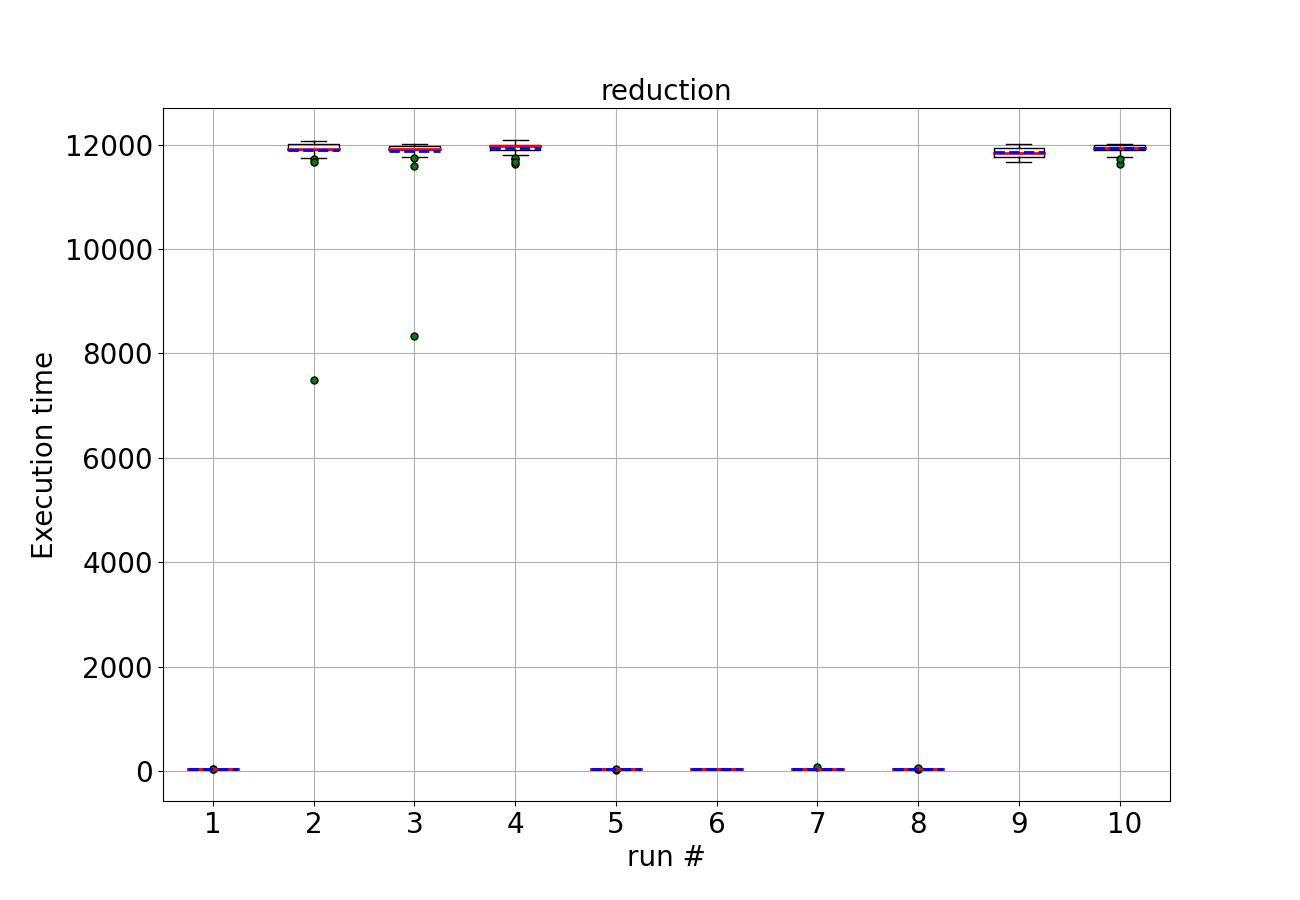}
\caption{\textit{syncbench} before thread-pinning}
\label{fig:dardel_reduction_wopin_exetime_outerrep_128core_st_8NUMA}}
\end{subfigure}
\begin{subfigure}[b]{0.3\linewidth}
{\includegraphics[width=1.1\textwidth]{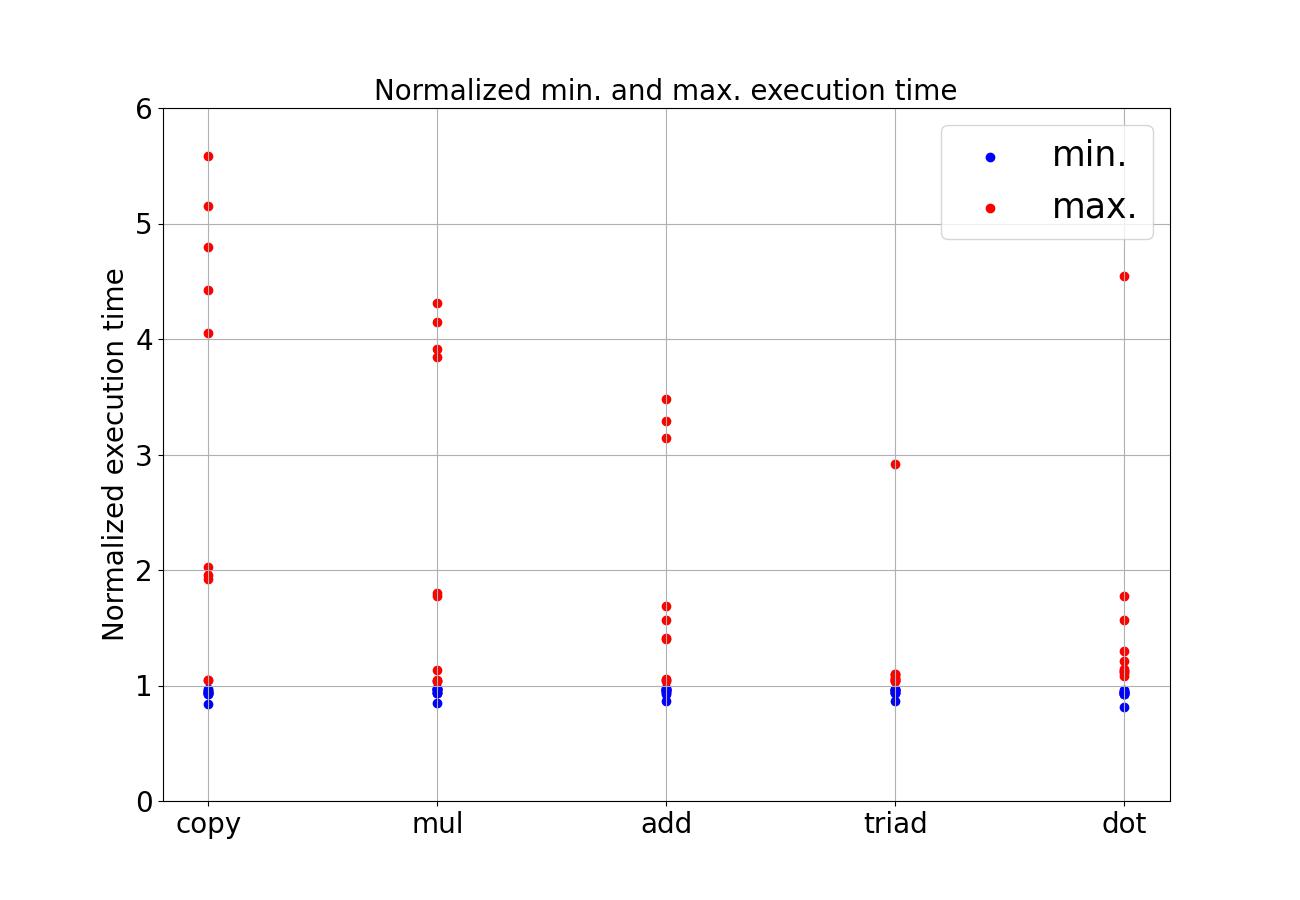}
\caption{\textit{BabelStream} before thread-pinning}
\label{fig:dardel_bstream_normarized_exetime_wopin_128core_st_8NUMA}}
\end{subfigure}
\quad
\begin{subfigure}[b]{0.3\linewidth}
{\includegraphics[width=1.0\textwidth]{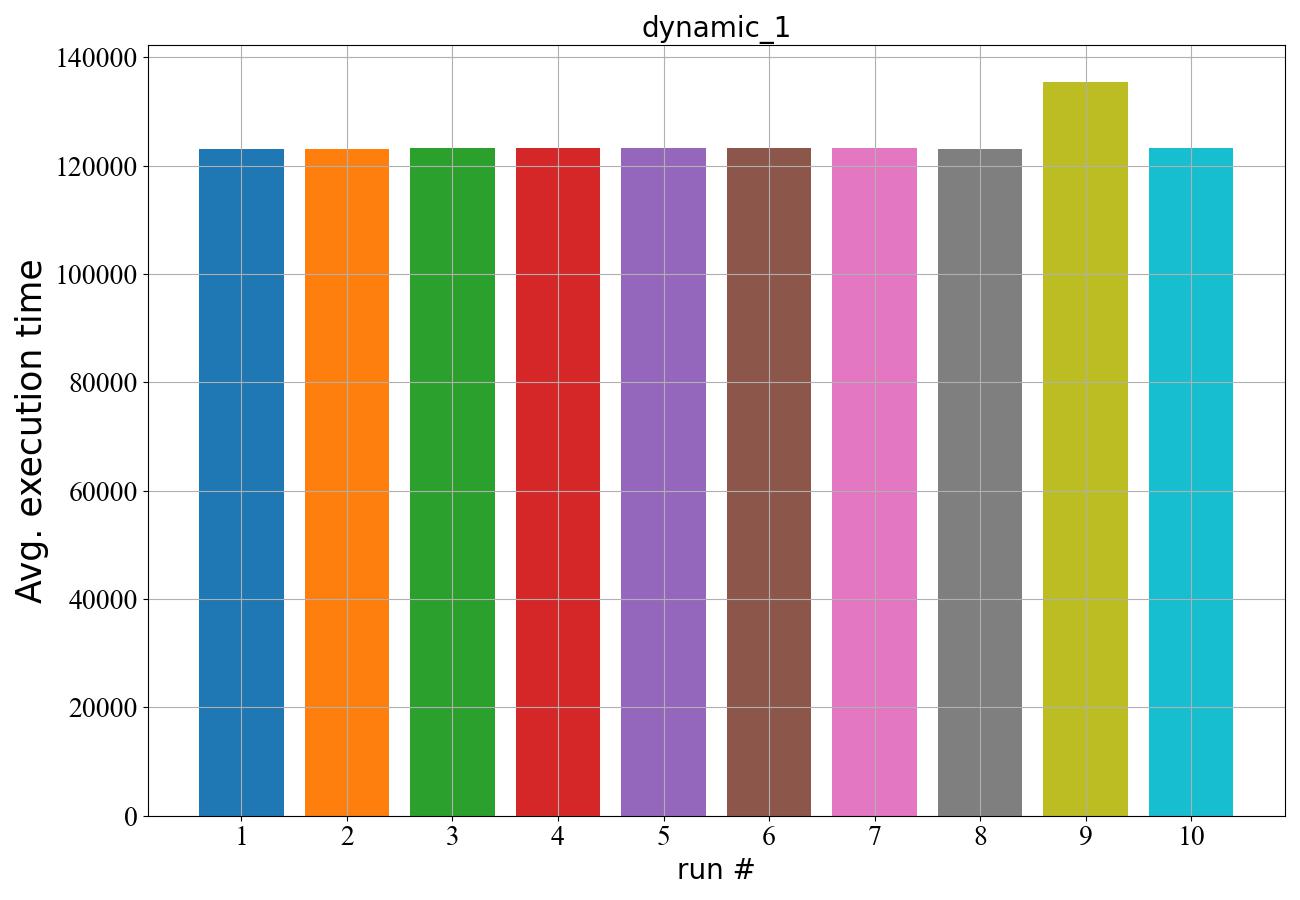}
\caption{\textit{schedbench} after thread-pinning}
\label{fig:dardel_dynamic_1_wpin_exetime_16core_st}}
\end{subfigure}
\begin{subfigure}[b]{0.3\linewidth}
{\includegraphics[width=1.1\textwidth]{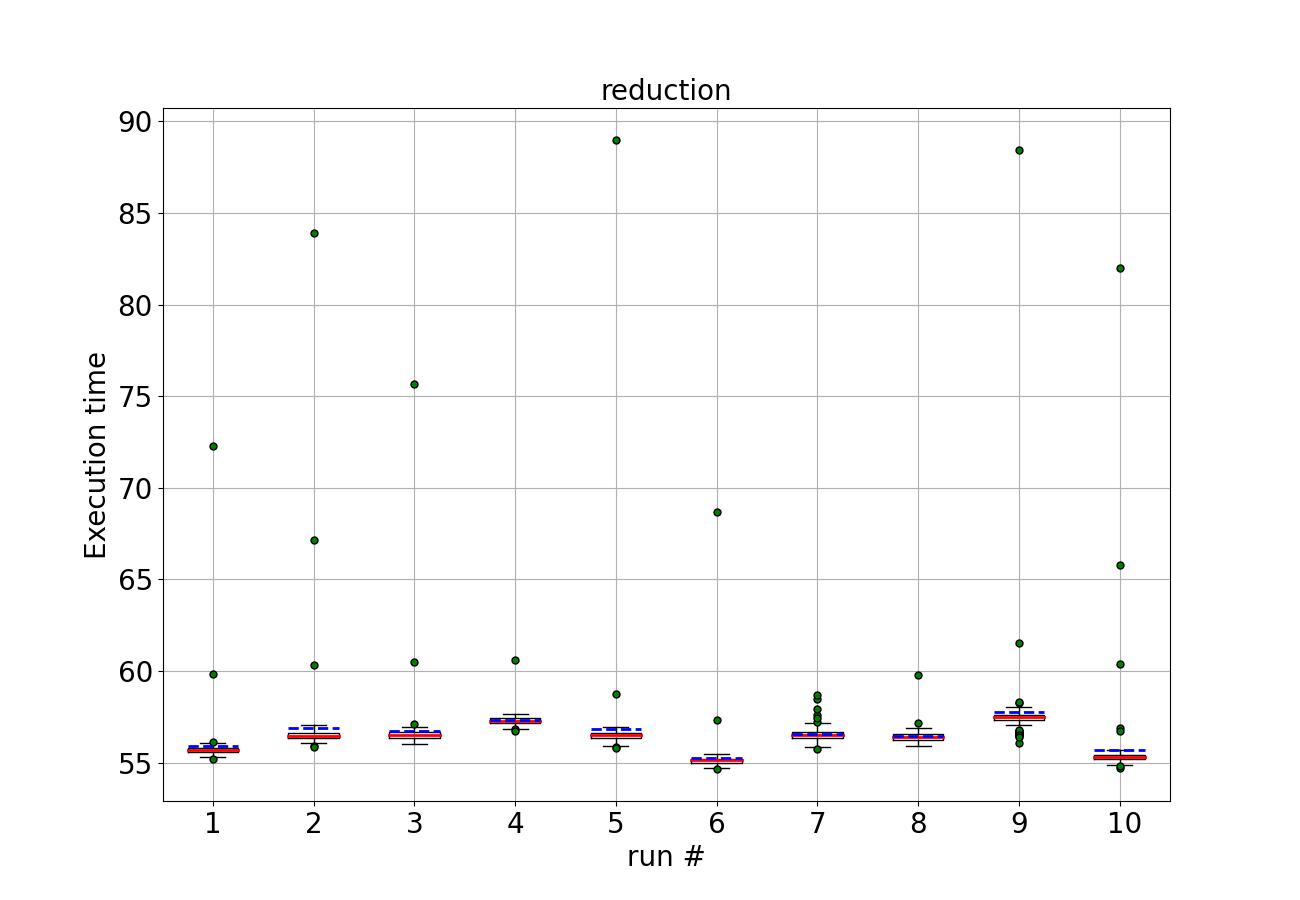}
\caption{\textit{syncbench} after thread-pinning}
\label{fig:dardel_reduction_wpin_exetime_outerrep_128core_st_8NUMA}}
\end{subfigure}
\begin{subfigure}[b]{0.3\linewidth}
{\includegraphics[width=1.1\textwidth]{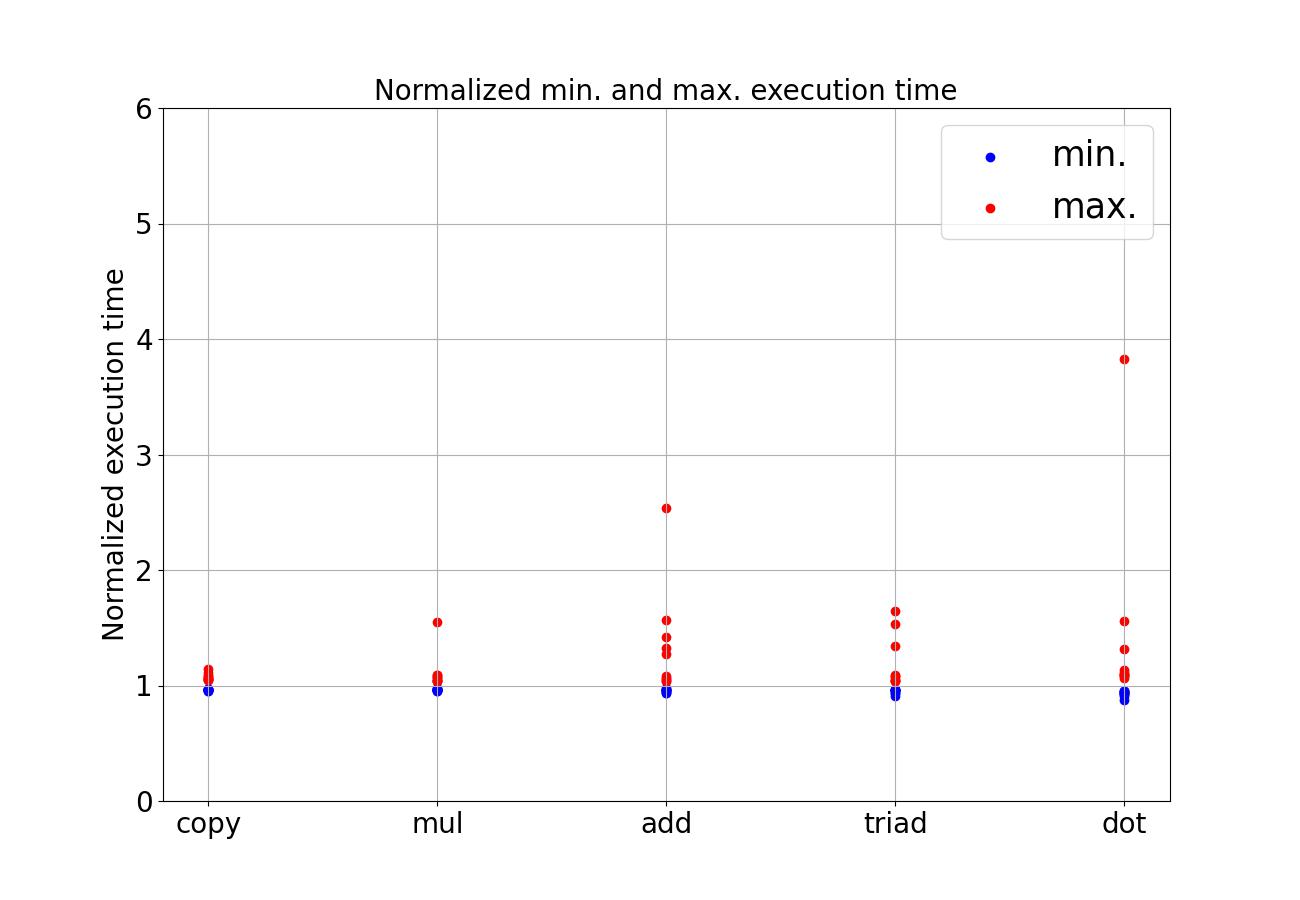}
\caption{\textit{BabelStream} after thread-pinning}
\label{fig:dardel_bstream_normarized_exetime_wpin_128core_st_8NUMA}}
\end{subfigure}
\caption{Lower variability of execution time(\textmu s) after thread-pinning in \textit{schedbench} (first column) when using 16 threads, and in \textit{syncbench} (second column) and in \textit{BabelStream} (third column) when using 128 threads on Dardel. }
\label{fig:lower_variability_after_thread_pin}
\vspace{-10pt}
\end{figure*}

In this part, we evaluate the effect of thread pinning on reducing performance variability. We showcase our results from Dardel in Figure~\ref{fig:lower_variability_after_thread_pin}, which is in accordance with our evaluation on Vera, although due to the larger scale of Dardel nodes, performance variability is more pronounced on this platform. 
The first column of Figure~\ref{fig:lower_variability_after_thread_pin} shows average (Avg.) execution times of \textit{schedbench} for 10 runs. The run-to-run variations of the average execution time do not completely disappear but can be reduced after the threads are pinned in Figure~\ref{fig:dardel_dynamic_1_wpin_exetime_16core_st}, where only run \# 9 has a higher execution time, compared to Figure~\ref{fig:dardel_dynamic_1_wopin_exetime_16core_st}, where runs \#(2,8,9,10) take longer time to finish. 
We believe that thread-pinning plays an important role in removing run-to-run variability and improving performance stability. \looseness=-1
The benefits of thread pinning are much more pronounced in the case of \textit{syncbench}, as synchronization primitives are more susceptible to noise and even slight increases in the execution time of one thread can propagate throughout the operation. Figure~\ref{fig:dardel_reduction_wopin_exetime_outerrep_128core_st_8NUMA} shows a high run-to-run variability for the reduction micro-benchmark of \textit{syncbench}, on 128 physical cores of Dardel, resulting in more than 3 
orders of magnitude of differences in the execution time of the micro-benchmark. Contrarily, after pinning, we achieve a much higher performance stability for the micro-benchmark, as shown in Figure~\ref{fig:dardel_reduction_wpin_exetime_outerrep_128core_st_8NUMA}. We note that the y-axes of the two subfigures have different scales. The run-to-run variability is almost eliminated after pinning, while the execution time variations between the 100 repetitions of the benchmark are also largely reduced for certain runs, e.g. runs \#(2, 3). 
Our observations for \textit{BabelStream}  are similar. Figure~\ref{fig:dardel_bstream_normarized_exetime_wopin_128core_st_8NUMA} shows high run-to-run variability for all the five kernels of the benchmark before thread pinning, as there is a difference of up to 6$\times$ between the minimum and maximum execution times between 10 different runs. After pinning, in Figure~
\ref{fig:dardel_bstream_normarized_exetime_wpin_128core_st_8NUMA}, we observe less run-to-run variability, especially for the \texttt{copy} and \texttt{mul} kernels. \looseness=-1

As load balancing at the OS-level can be affected by thread-pinning, it is promising to jointly consider pinning policy and application characteristics. In a nutshell, thread pinning is particularly beneficial for reducing run-to-run variability and improving performance stability of OpenMP applications, especially for memory-bound applications, as evidenced by \textit{BabelStream} and synchronization-sensitive applications, as evidenced by \textit{syncbench}. In the remainder of our evaluation, we use thread pinning for all our experiments.  \looseness=-1

\begin{figure*}[htbp]
\centering

\begin{subfigure}[b]{0.3\textwidth}
{\includegraphics[width=1.1\textwidth]{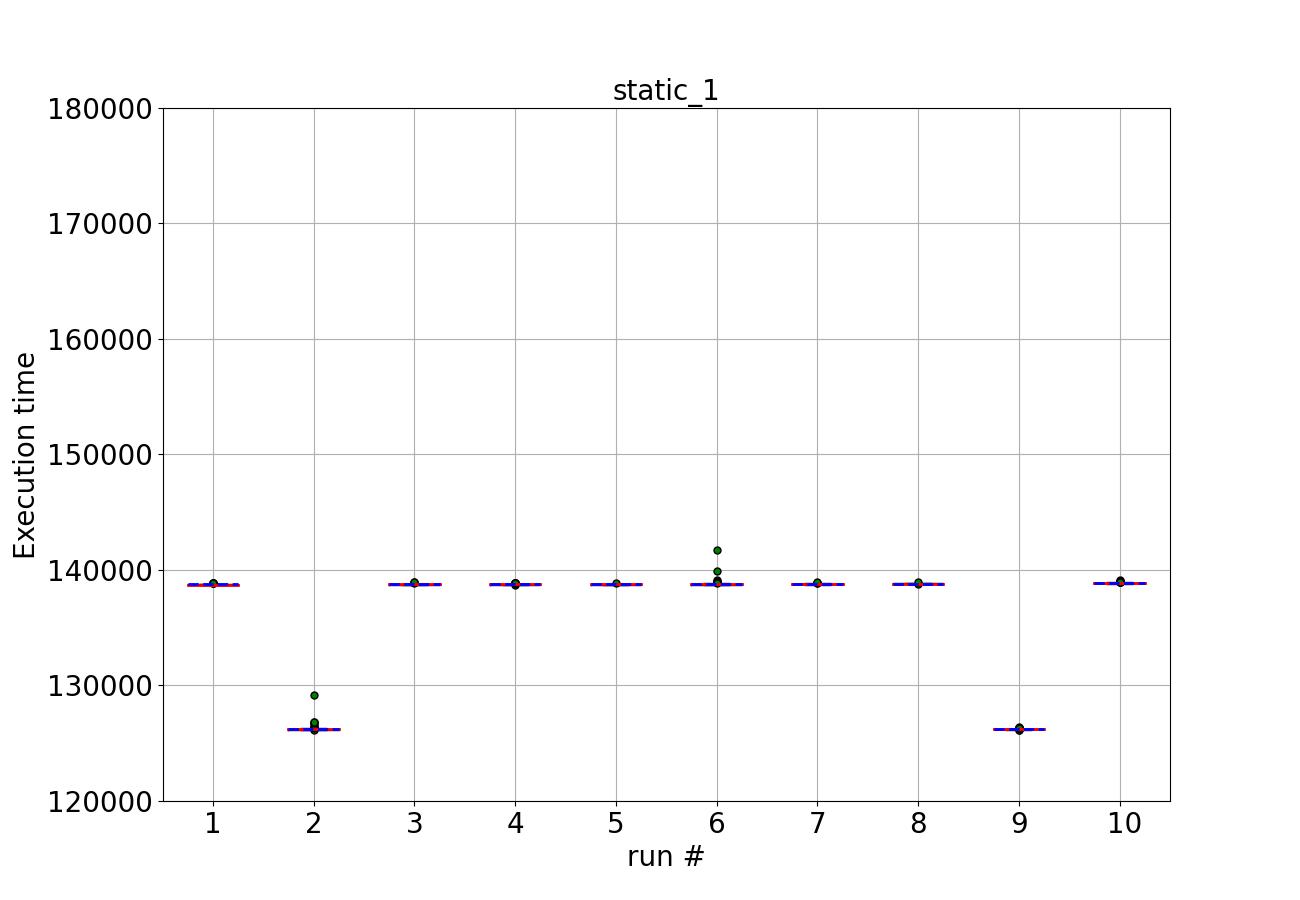}
\caption{\textit{schedbench} with ST}
\label{fig:dardel_sched_static1_exetime_10runs_128core_st_NUMA_new}}
\end{subfigure}
\begin{subfigure}[b]{0.3\textwidth}
{\includegraphics[width=1.1\textwidth]{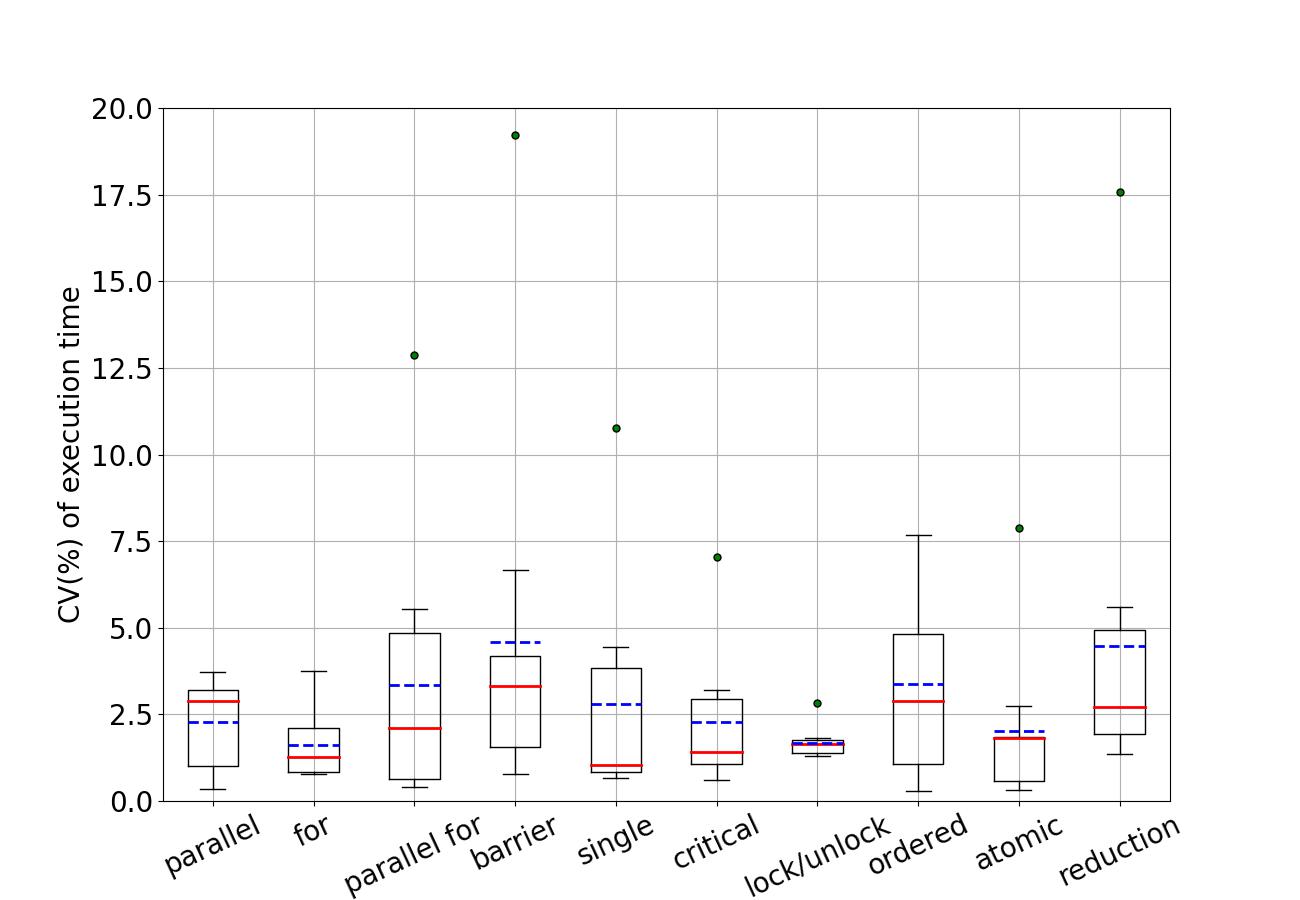}
\caption{\textit{syncbench} with ST }
\label{fig:dardel_sync_cv_10runs_32thread_st_NUMA_new}}
\end{subfigure}
\begin{subfigure}[b]{0.3\textwidth}
{\includegraphics[width=1.1\textwidth]{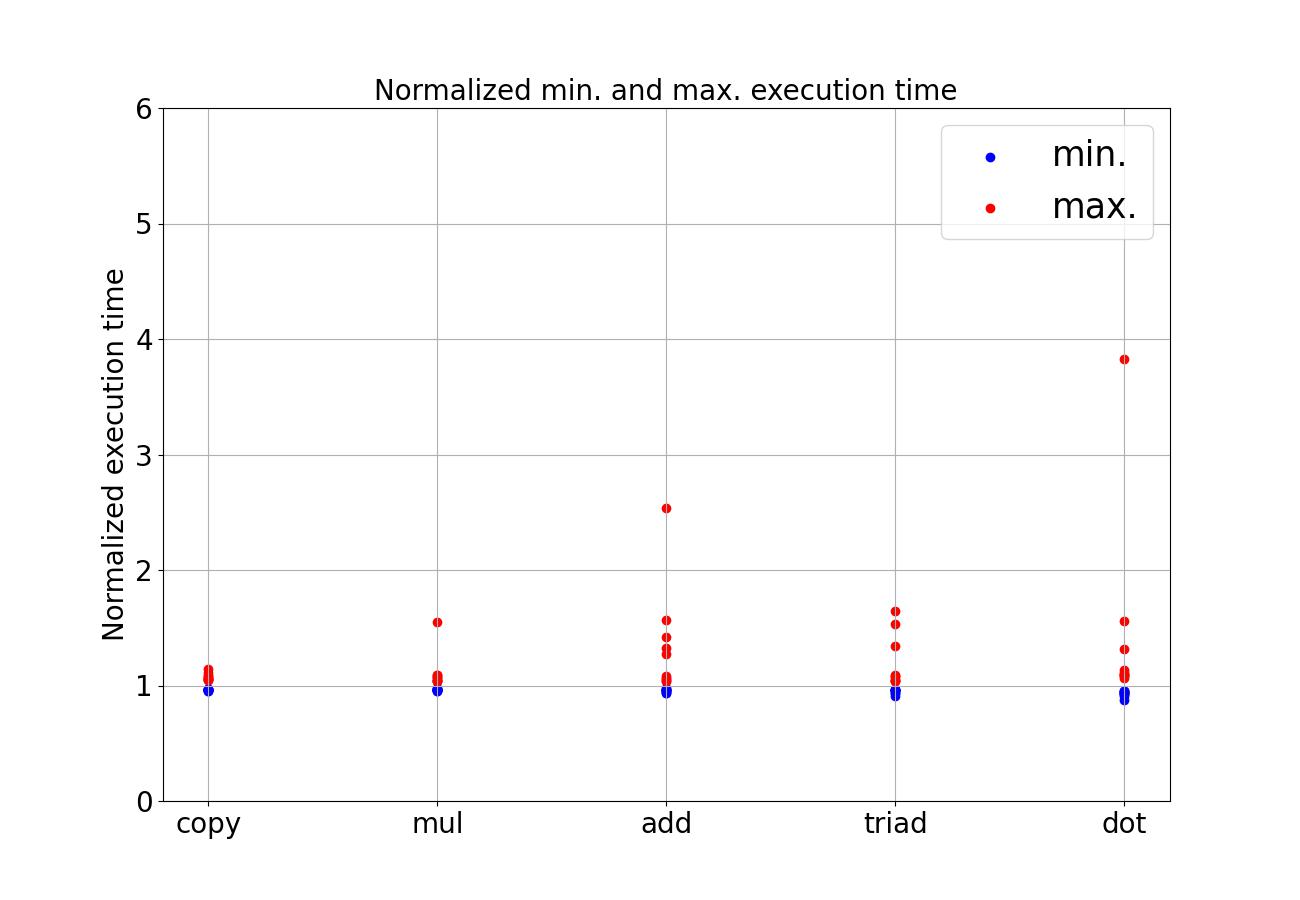}
\caption{\textit{BabelStream} with ST}
\label{fig:dardel_bstream_normarized_min_maxexetime_128thread_ST_8NUMA_new}}
\end{subfigure}
\quad
\begin{subfigure}[b]{0.3\textwidth}
\centering
{\includegraphics[width=1.1\textwidth]{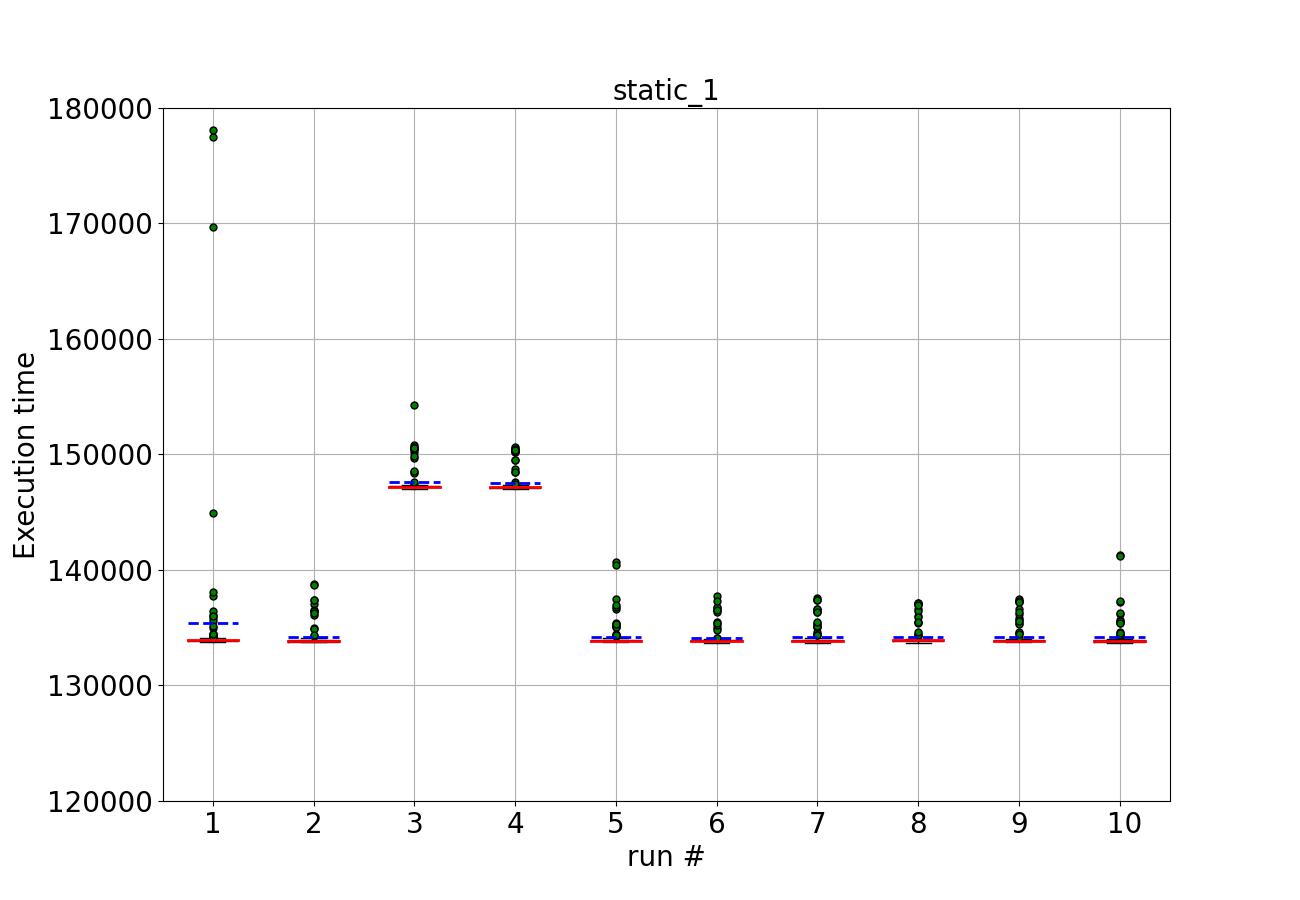}
\caption{\textit{schedbench} with MT}
\label{fig:dardel_sched_static1_exetime_10runs_128core_mt_NUMA_new}}
\end{subfigure}
\begin{subfigure}[b]{0.3\textwidth}
\centering
{\includegraphics[width=1.1\textwidth]{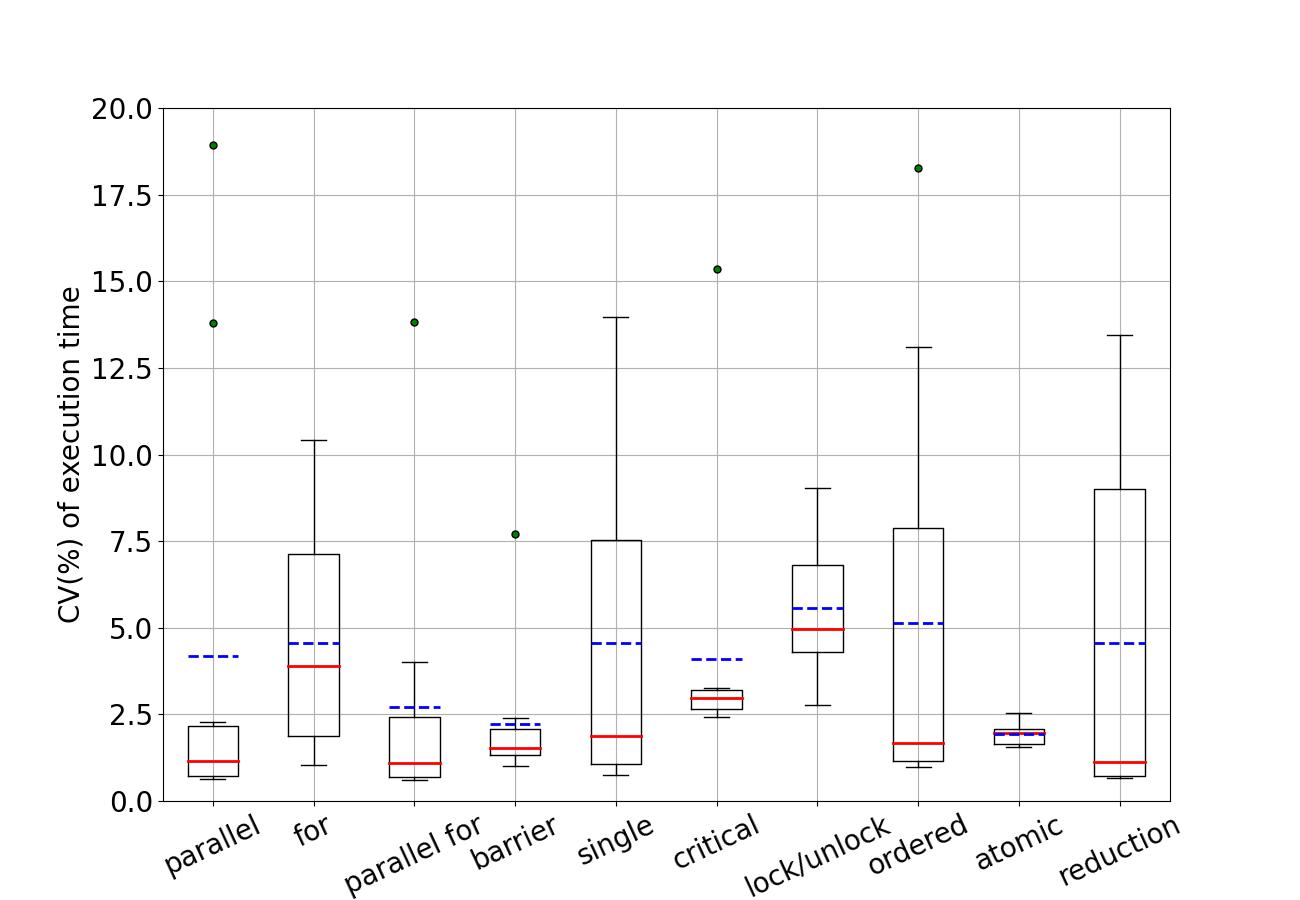}
\caption{\textit{syncbench} with MT}
\label{fig:dardel_sync_cv_10runs_32thread_mt_NUMA_new}}
\end{subfigure}
\begin{subfigure}[b]{0.3\textwidth}
{\includegraphics[width=1.1\textwidth]{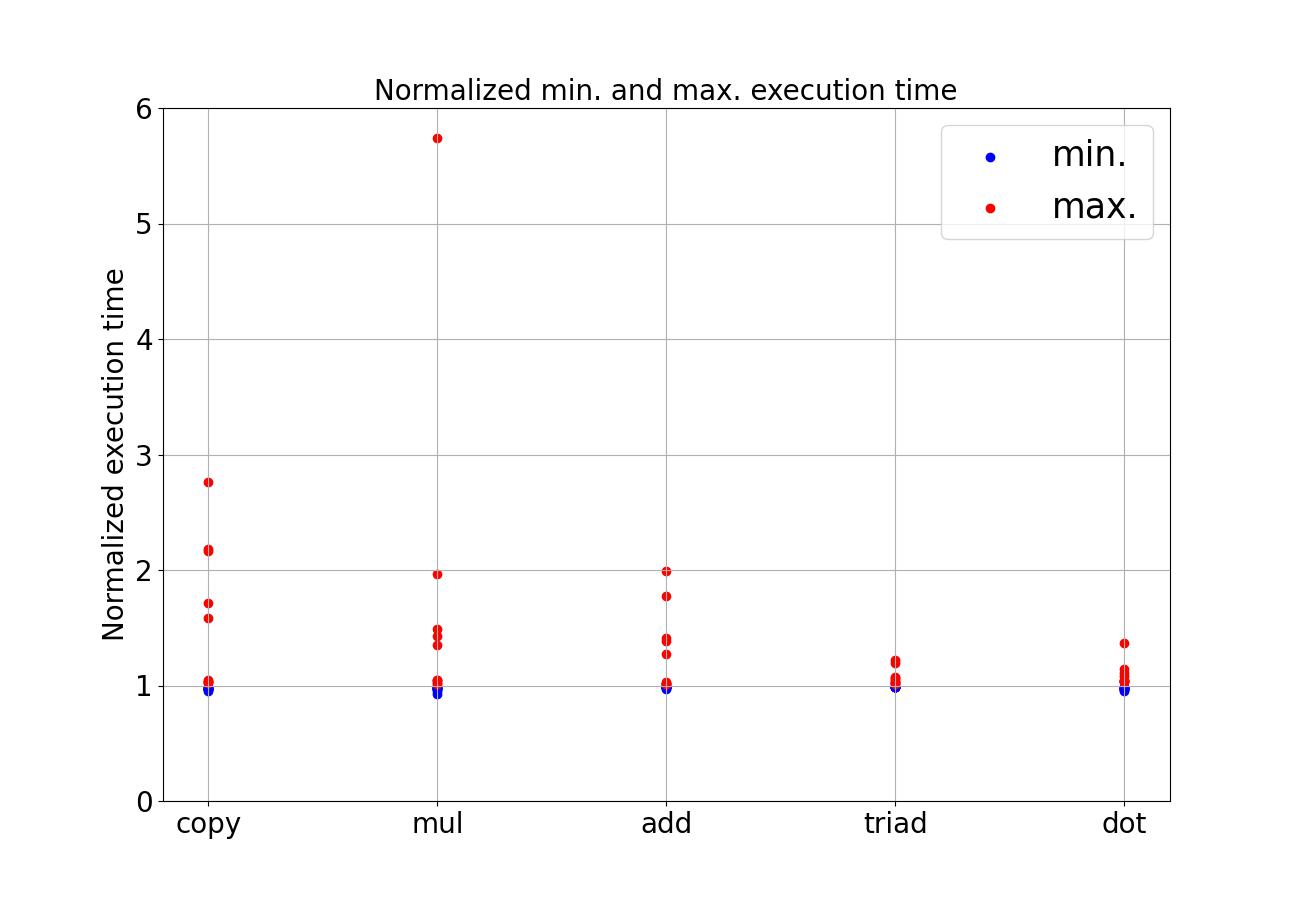}
\caption{\textit{BabelStream} with MT}
\label{fig:dardel_bstream_normarized_min_maxexetime_128thread_MT_8NUMA_new}}
\end{subfigure}
\caption{Higher variability of execution time (\textmu s) in \textit{schedbench} ( first column, executed with 128 threads) and \textit{syncbench} (second column, executed with 32 threads) and \textit{BabelStream} (third column, executed with 128 threads) due to SMT implementation on Dardel.}
\label{fig:performance _variability_dardel_ST_SMT}
\vspace{-10pt}
\end{figure*}

\subsection{The effect of SMT}
We examine the effect of simultaneous multithreading on Dardel, as Vera does not support SMT. We compare the performance variability of our benchmarks in Figure~\ref{fig:performance _variability_dardel_ST_SMT}, for the \textbf{ST} 
case, where we use only physical cores of Dardel, e.g.. 32/64/128 cores and OpenMP threads, and the \textbf{MT} case, where we use both two hardware threads of 16/32/64 physical cores of Dardel, i.e. 32/64/128 HW threads and OpenMP threads. 
We note that the use of SMT is usually decided by the developer/user, based on application properties, e.g. the compute-boundedness or memory-boundedness of the application. However, in our evaluation, we regard SMT only as a potential source of performance variability, examining cases where we use the same number of threads. 


For \textit{schedbench} in Figure~\ref{fig:dardel_sched_static1_exetime_10runs_128core_st_NUMA_new} and Figure~\ref{fig:dardel_sched_static1_exetime_10runs_128core_mt_NUMA_new}, even though some run-to-run variability exists under the \textbf{ST} configuration, we observe a very high variability among the 100 outer repetitions of the benchmark, for each single run under the \textbf{MT} configuration.
Regarding \textit{syncbench},  
we compare the run-to-run variations of execution times and adopt the coefficient of variation (CV), i.e. the ratio of the standard deviation to the average (lower is better), for every run, as the metric to measure the performance variability of execution time in Figure~\ref{fig:dardel_sync_cv_10runs_32thread_st_NUMA_new} and Figure~\ref{fig:dardel_sync_cv_10runs_32thread_mt_NUMA_new}. The performance stability is significantly affected in a negative way when leveraging SMT especially for some synchronization directives such as \texttt{for},\texttt{single},\texttt{ordered} and \texttt{reduction}, as the CV values of all 10 runs show high variances in Figure~\ref{fig:dardel_sync_cv_10runs_32thread_mt_NUMA_new}. 
For most synchronization cases, the \textbf{ST} configuration exhibits better performance stability in Figure~\ref{fig:dardel_sync_cv_10runs_32thread_st_NUMA_new}, by leaving the second hardware thread free, potentially available for OS activities, whereas higher performance variability, including run-to-run variations and the variations among the 100 outer repetitions for each single run can be seen with the \textbf{MT} configuration.
Similarly, we compare the performance variability of the normalized minimum and maximum times for all 10 runs for \textit{BabelStream} in Figure~\ref{fig:dardel_bstream_normarized_min_maxexetime_128thread_ST_8NUMA_new} and Figure~\ref{fig:dardel_bstream_normarized_min_maxexetime_128thread_MT_8NUMA_new} under the \textbf{ST} and \textbf{MT} configurations respectively. \textit{BabelStream} also does not benefit from using hardware threads.

The above observations reveal that leaving the second thread in SMT implementation for system activities results in better performance stability, while the \textbf{MT} configuration makes the executing benchmark experience more SMT interference. This impact of additional hardware thread resources reserved for operating system, i.e. \textbf{ST} configuration, varies with benchmark characteristics and scale. For example, \textbf{ST} does not outperform \textbf{MT} much for BabelStream when only a few threads are used. Overall, leaving the additional thread resources implemented by SMT mechanism for OS activities can be a promising way to achieve performance stability for the OpenMP runtime. 

\subsection{The effect of frequency variation}

\begin{figure}[tbp]

\centering
\begin{subfigure}[b]{0.4\textwidth}
\includegraphics[width=\textwidth]{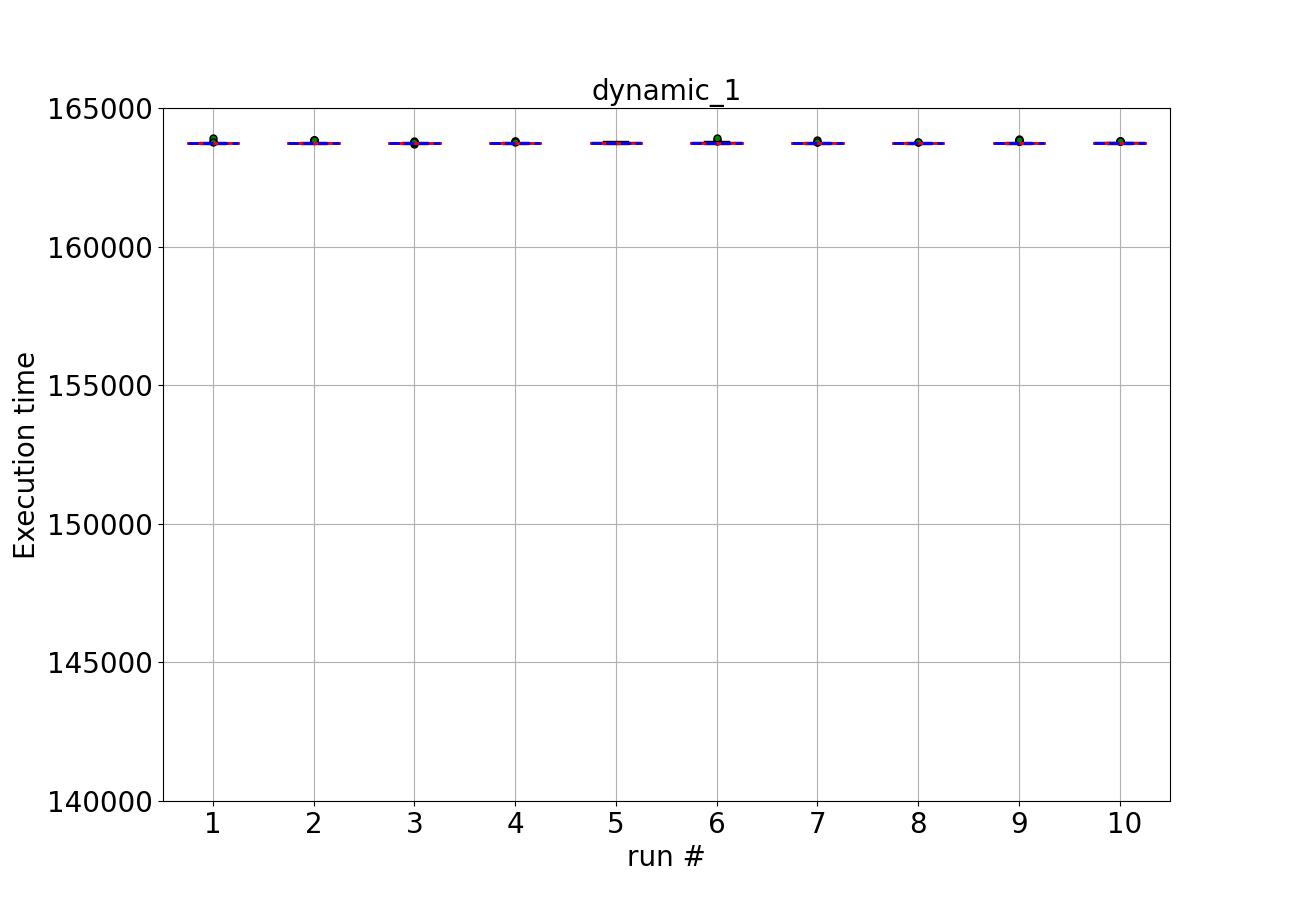}
\caption{16 cores from one NUMA node}
\label{fig:vera_dynamic1_exetime_outerrep_16core_st}
\end{subfigure}
\begin{subfigure}[b]{0.4\textwidth}
\includegraphics[width=\textwidth]{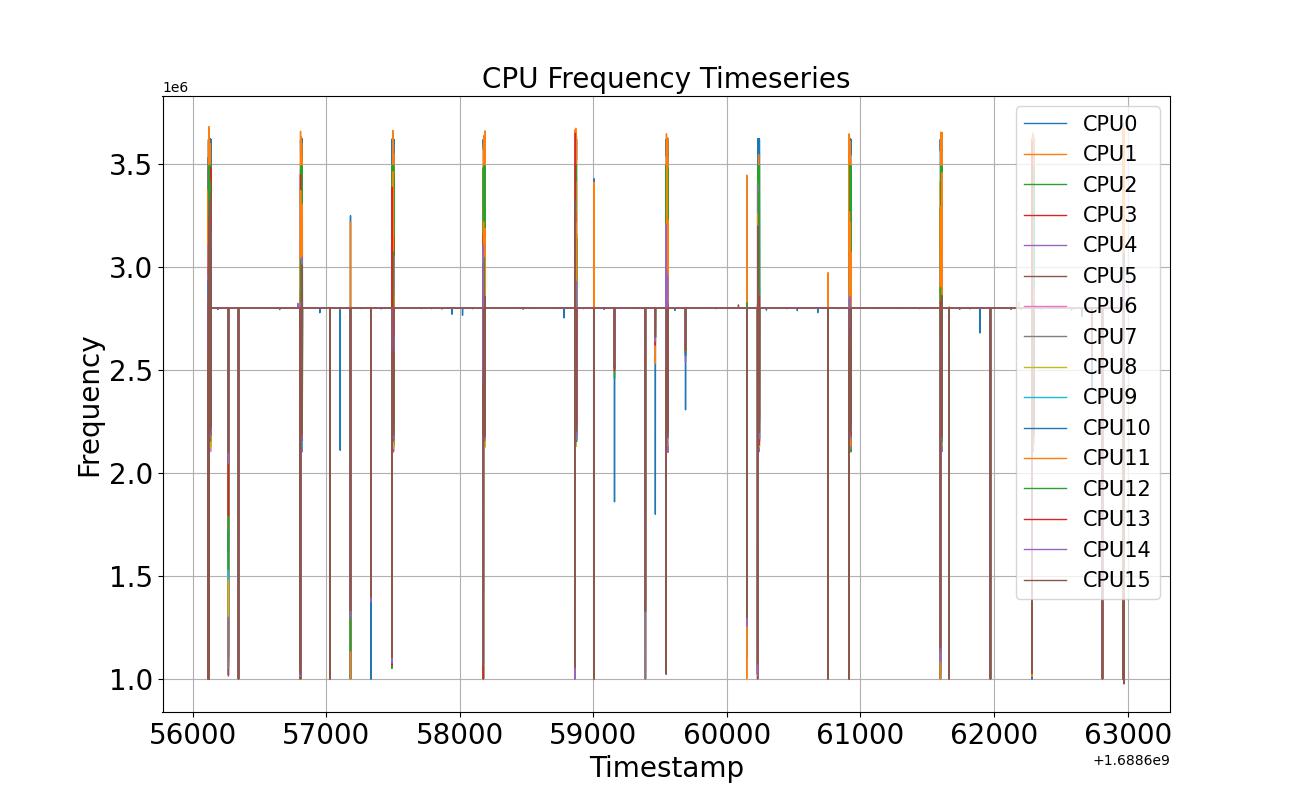}
\caption{frequency}
\label{fig:vera_sched_freq_16core_st_10times}
\end{subfigure}
\begin{subfigure}[b]{0.4\textwidth}
\includegraphics[width=\textwidth]{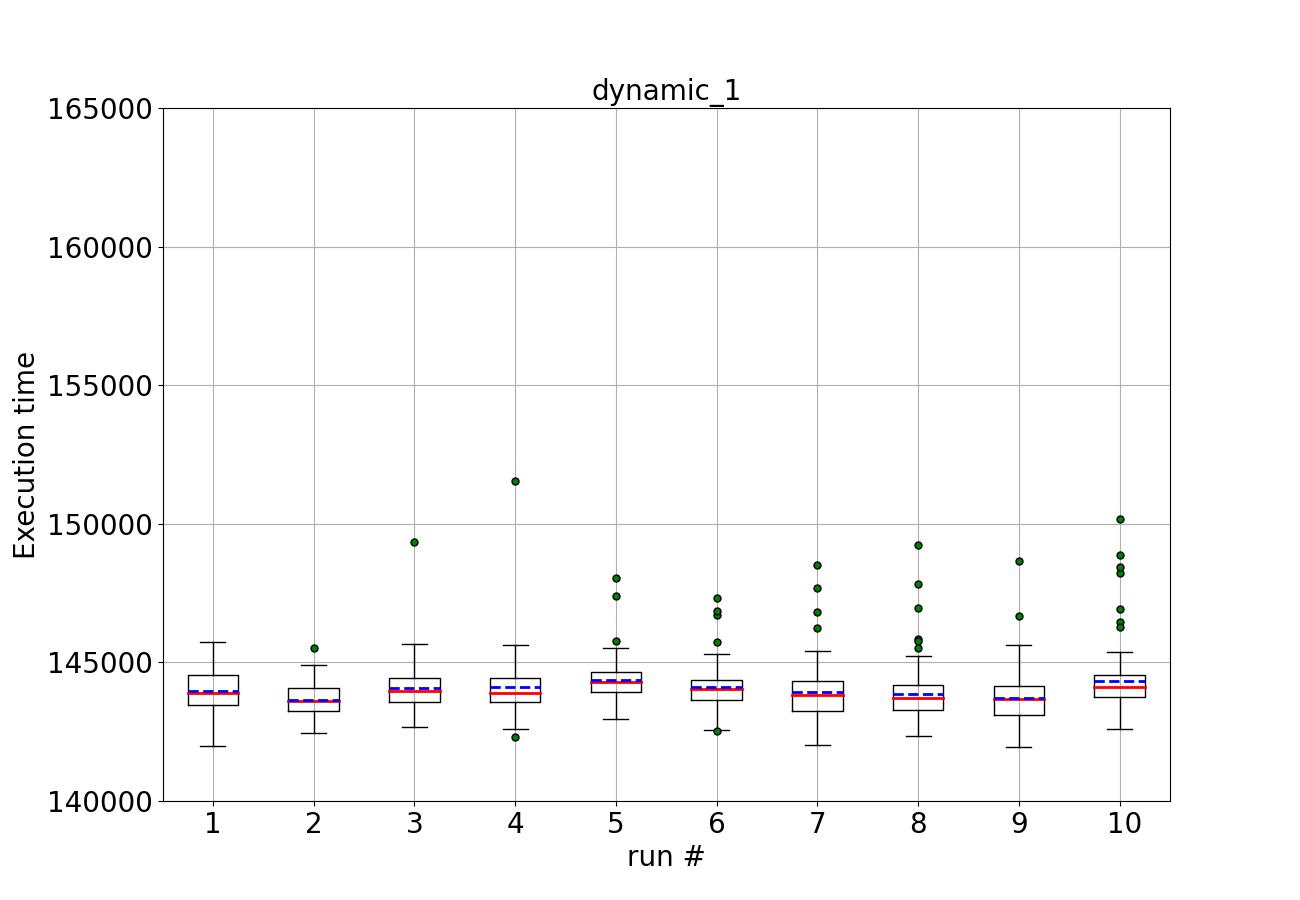}
\caption{16 cores from two NUMA nodes}
\label{fig:vera_dynamic1_exetime_outerrep_16core_st_NUMA}
\end{subfigure}
\begin{subfigure}[b]{0.4\textwidth}
\includegraphics[width=\textwidth]{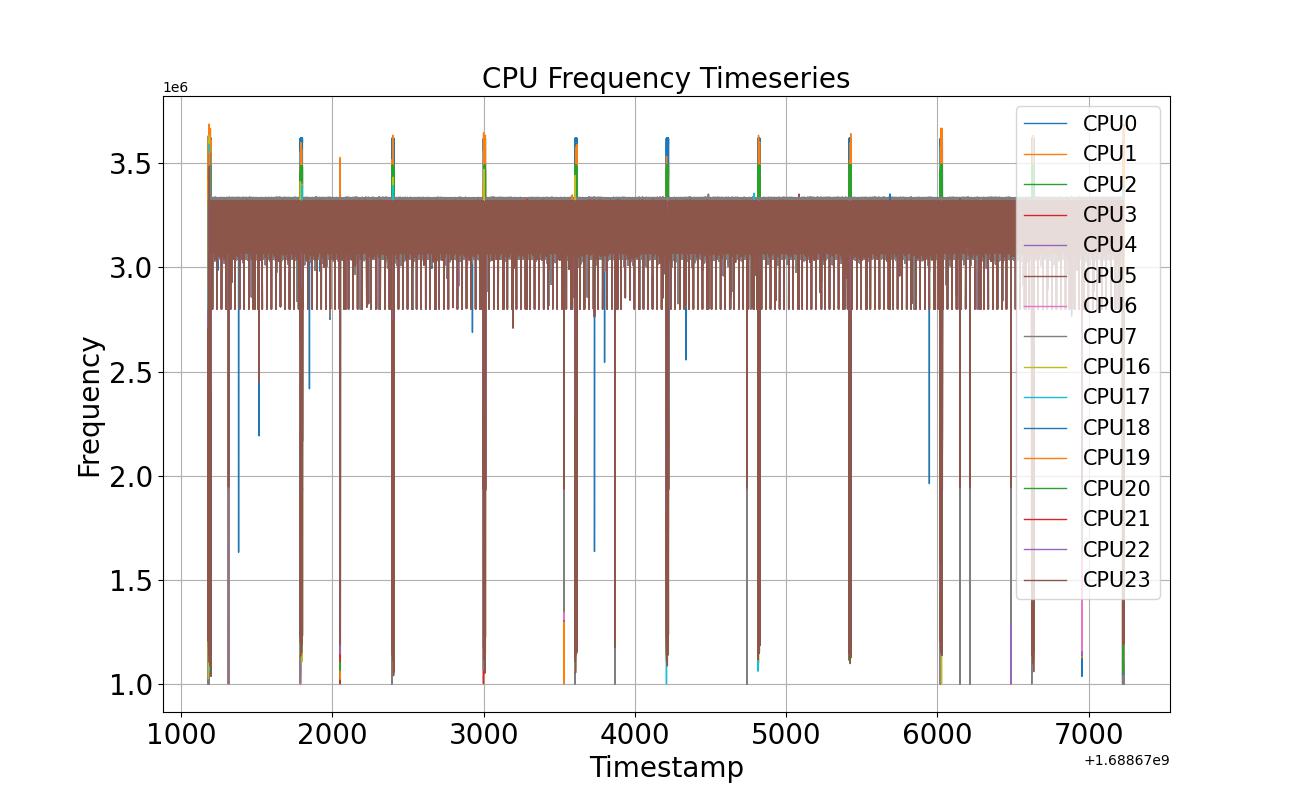}
\caption{frequency}
\label{fig:vera_sched_freq_16core_st_NUMA_10times}
\end{subfigure}
\caption{Higher variability of execution time(\textmu s) in \textit{schedbench} due to frequency variation on Vera. }
\label{fig:variability_freq_shaking_schedbench}
\end{figure}

\begin{figure}[tbp]

\centering
\begin{subfigure}[b]{0.4\textwidth}
\centering
\includegraphics[width=\textwidth]{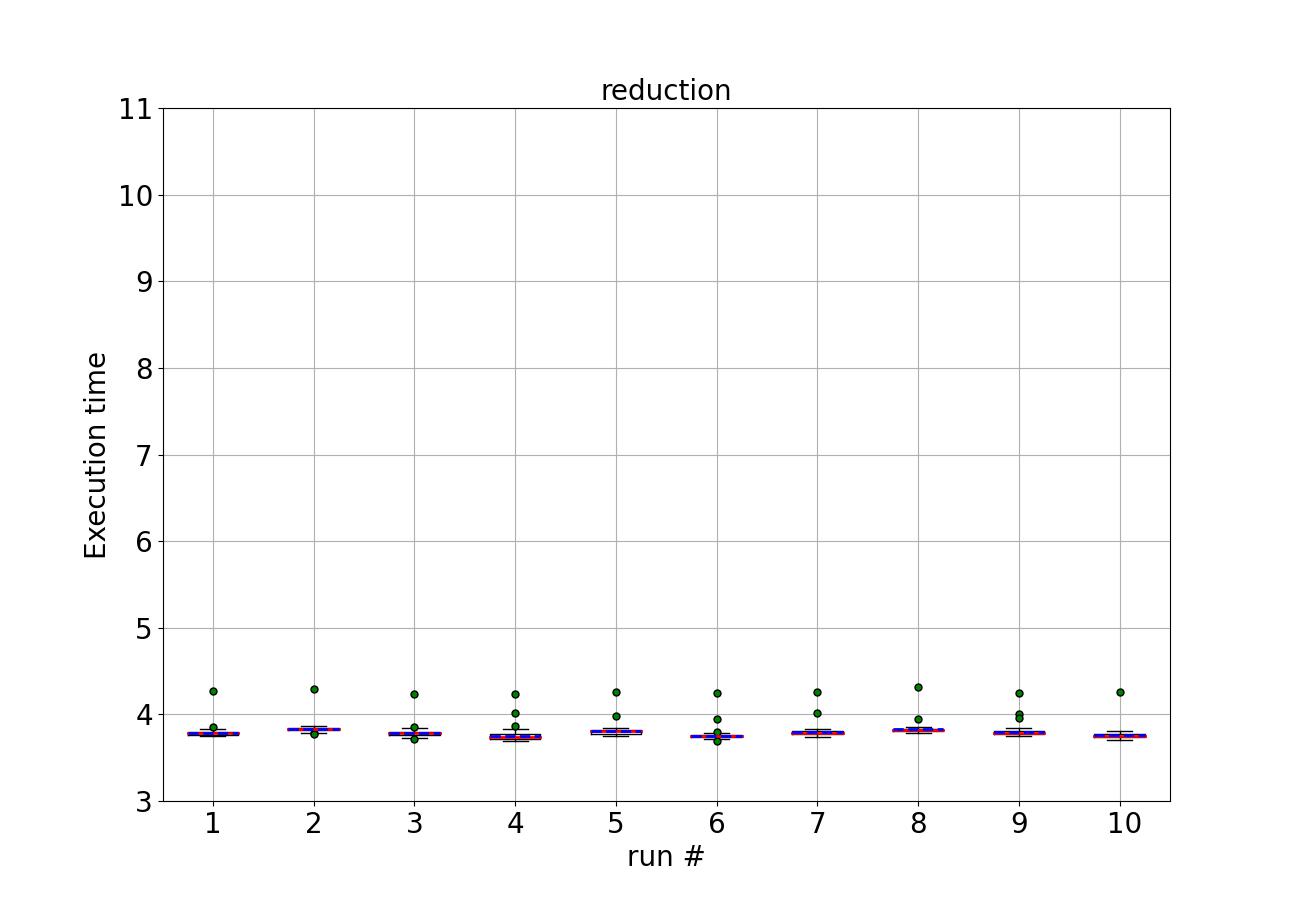}
\caption{16 cores from one NUMA node}
\label{fig:vera_reduction_exetime_outerrep_16core_st}
\end{subfigure}
\begin{subfigure}[b]{0.4\textwidth}
\centering
\includegraphics[width=\textwidth]{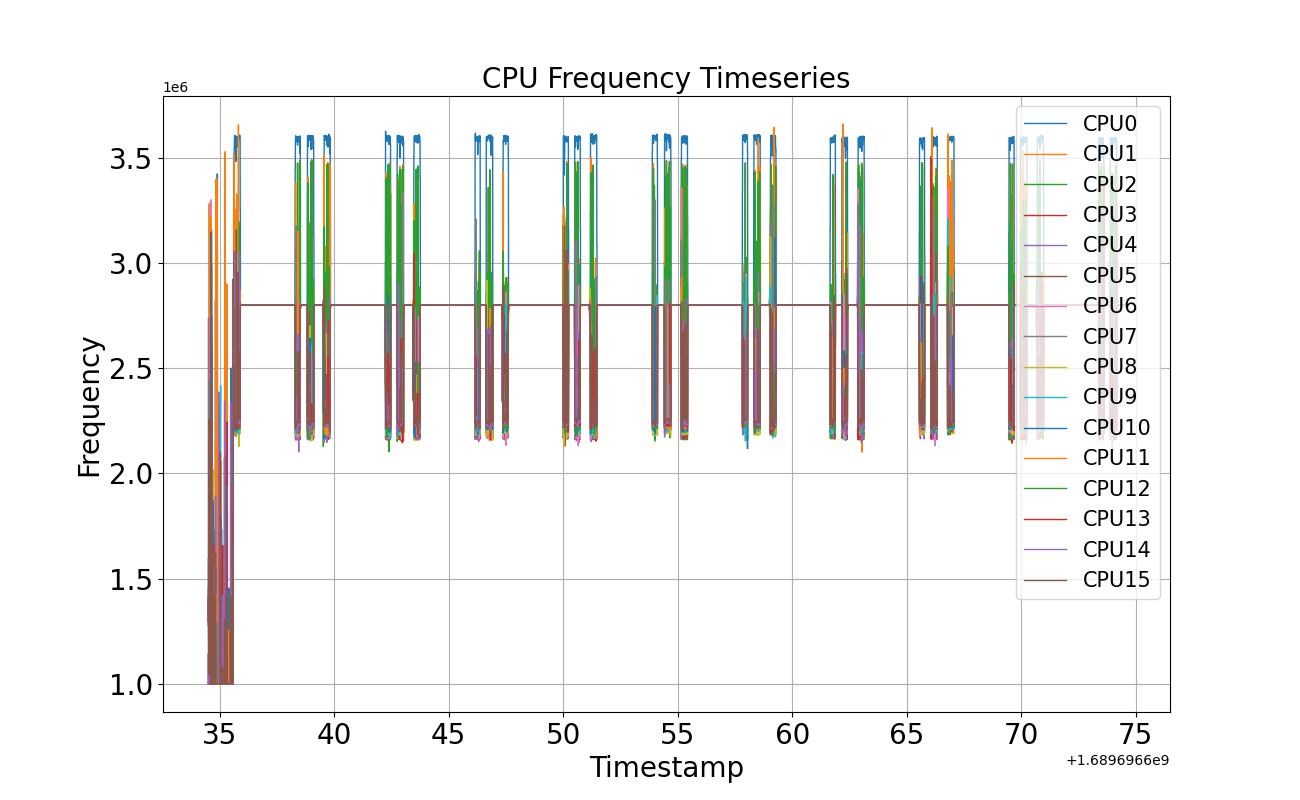}
\caption{frequency}
\label{fig:vera_sync_freq_16core_st_10times}
\end{subfigure}
\begin{subfigure}[b]{0.4\textwidth}
\centering
\includegraphics[width=\textwidth]{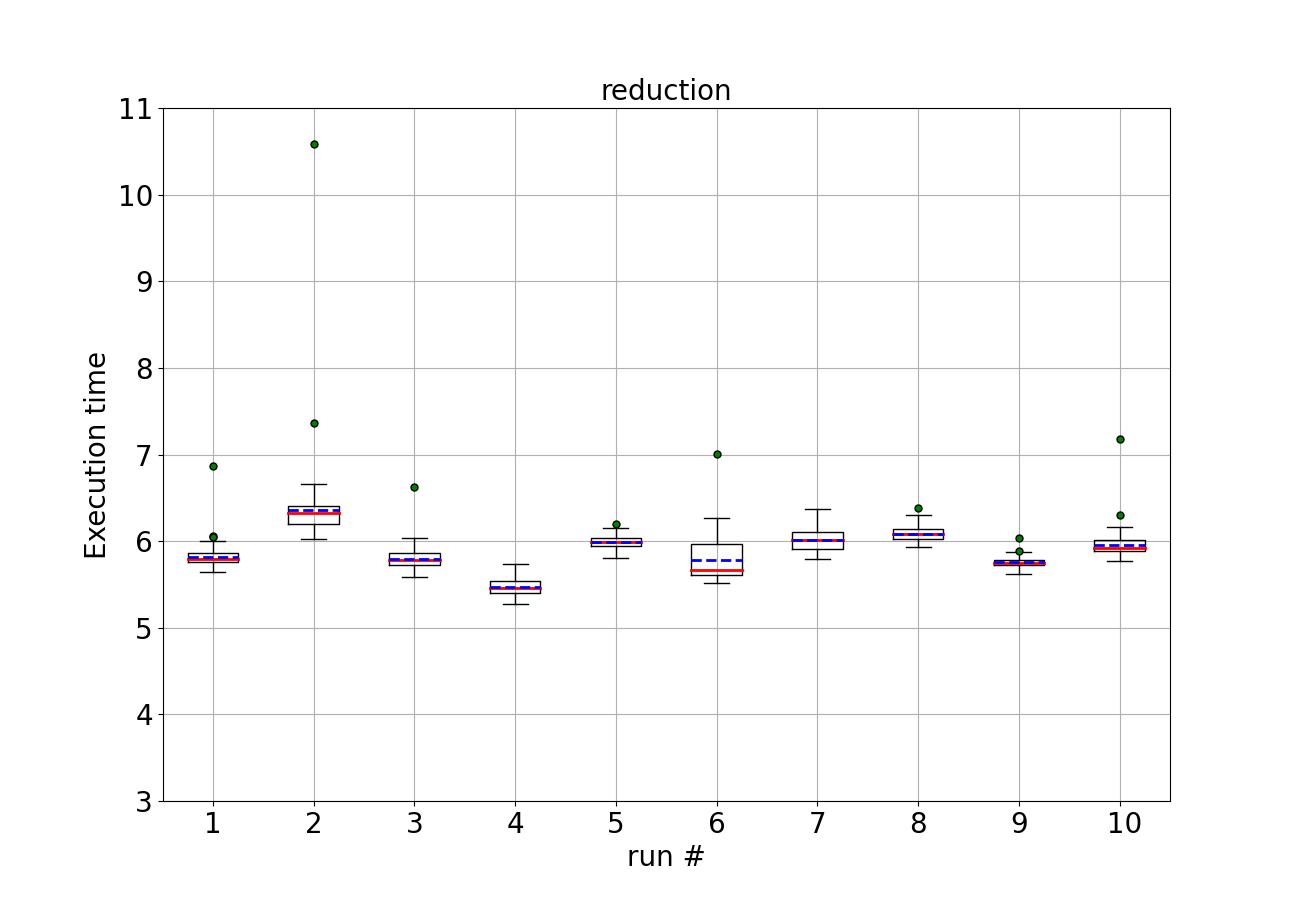}
\caption{16 cores from two NUMA nodes}
\label{fig:vera_reduction_exetime_outerrep_16core_st_NUMA}
\end{subfigure}
\begin{subfigure}[b]{0.4\textwidth}
\centering
\includegraphics[width=\textwidth]{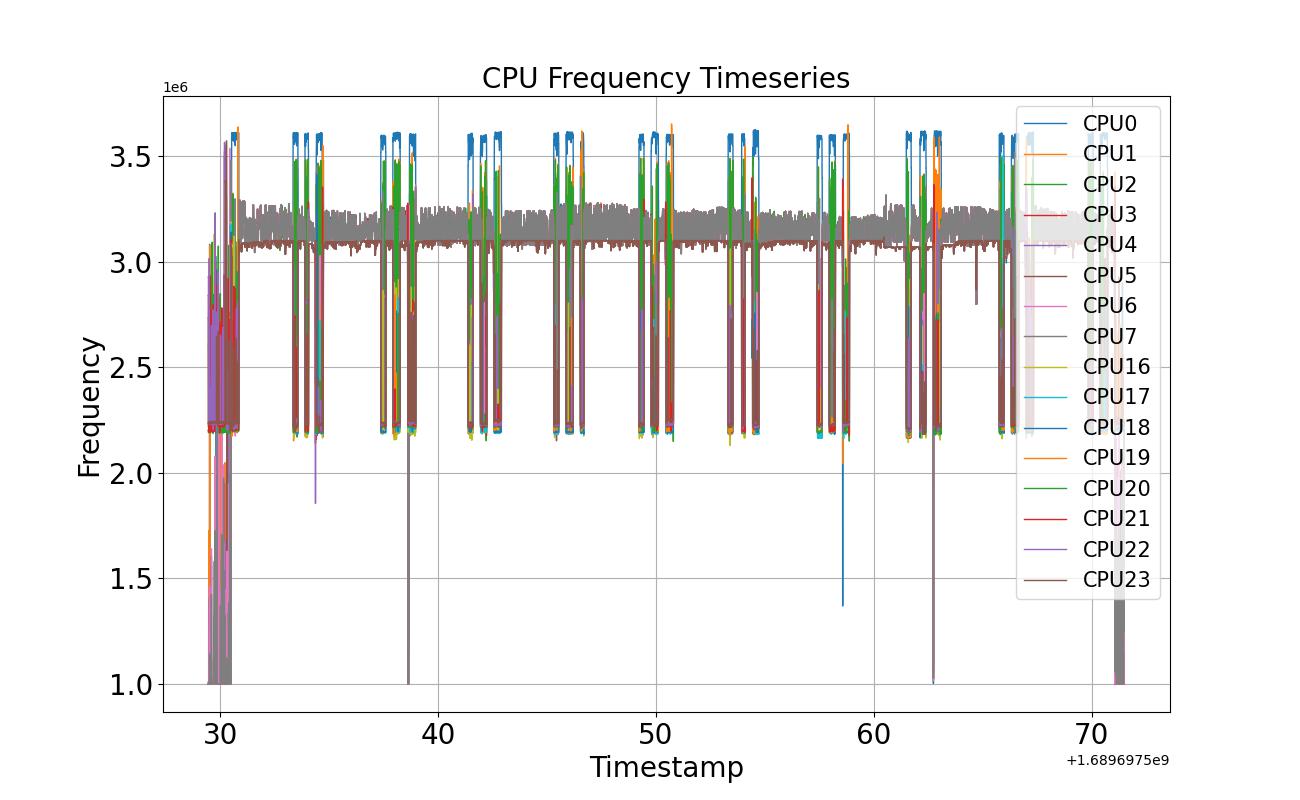}
\caption{frequency}
\label{fig:vera_sync_freq_16core_st_NUMA_10times}
\end{subfigure}
\caption{Higher execution time(\textmu s) in \textit{syncbench} micro-benchmark due to frequency variation on Vera.}
\label{fig:variability_freq_shaking_syncbench}
\end{figure}

We finally examine the effect of frequency variation on the performance variability of OpenMP, by continuously logging the frequency levels of all cores (read through the sysfs interface of the Linux CPUFreq), through a Python script executing on a separate core. Although the default governor on Vera is set to performance, boosting all the core frequencies to the maximum, for some of our experiments, we have observed performance variability, especially across NUMA nodes, which can be justified by frequency variations. 
For some of the experiments done on Vera when using same cores but from same NUMA node or cross NUMA nodes, we observed the different behaviours of performance variability that can be potentially explained by the variation of frequency.
Figure~\ref{fig:variability_freq_shaking_schedbench} depicts the relation between variability of execution time for \textit{schedbench} and frequency variation. Figure~\ref{fig:vera_dynamic1_exetime_outerrep_16core_st_NUMA}, where we use cores across NUMA nodes, shows higher performance variability, both between different runs, and among the 100 outer repetitions, compared to Figure~\ref{fig:vera_dynamic1_exetime_outerrep_16core_st}, where we use the same number of cores on a single NUMA node. Figure~\ref{fig:vera_sched_freq_16core_st_10times} and Figure~\ref{fig:vera_sched_freq_16core_st_NUMA_10times} depict the behavior of frequency for these two groups of experiments respectively. The brown region in Figure~\ref{fig:vera_sched_freq_16core_st_NUMA_10times} during all 10 runs indicates more frequent frequency variation, compared to Figure~\ref{fig:vera_sched_freq_16core_st_10times}. As higher frequency levels used to run the benchmark, dictated by the performance governor, positively influence the execution time, this explains the above observation that the execution times vary more, depending on the frequency variation. \looseness=-1

We made a similar observation for \textit{syncbench} in Figure~\ref{fig:variability_freq_shaking_syncbench}, where Figure~\ref{fig:vera_reduction_exetime_outerrep_16core_st_NUMA} exhibits more variations for both run-to-run executions and outer repetitions for a single run compared to Figure~\ref{fig:vera_reduction_exetime_outerrep_16core_st}. The same effect of frequency variation can be seen in the grey region in Figure~\ref{fig:vera_sync_freq_16core_st_NUMA_10times}.
%
We note that on Dardel, we have not observed an obvious trend between performance variability and frequency variations, as Dardel exhibits less frequency variation compared to Vera. 

\vspace*{-0.2cm}
\section{Conclusion}\label{conclusion}

This paper aims to characterize the performance variability of OpenMP benchmarks and analyze the potential sources and impact of the performance variability based on an experimental study. We have tested two OpenMP benchmarks from the EPCC OpenMP micro-benchmark suite and BabelStream,  on two platforms, assessing the impact of thread pinning, SMT, and core frequency variation on performance variability. 
Our experimental results have illustrated that performance variability exists in OpenMP, both within a benchmark and between different runs, and can be reduced considerably by applying thread-pinning, leaving the additional hardware threads implemented by SMT for OS activities, but can be negatively affected by frequency variation during execution, which is beyond the control of the user. 

For future work, we aim to extend our characterization to other benchmarks such as FP-intensive or cache-intensive benchmarks and larger OpenMP applications on other platforms. We also wish to pinpoint the exact sources of OS noise and their impact on OpenMP applications, in order to design strategies to mitigate or eliminate performance variability. 


\begin{acks}
The experiments were enabled by resources provided by the National Academic Infrastructure for Supercomputing in Sweden (NAISS) at PDC partially funded by the Swedish Research Council through grant agreement no. \mbox{2022-06725}. 
The experiments were furthermore enabled by resources provided by Chalmers e-Commons at Chalmers.
This project has received funding from the European High Performance Computing Joint Undertaking (JU) under Framework Partnership Agreement No.~800928 and Specific Grant Agreement No.~101036168 (EPI SGA2). The JU receives support from the European Union’s Horizon 2020 research and innovation programme, and from the Swedish Research Council, among others. 
\end{acks}

  \bibliographystyle{ACM-Reference-Format}
  \bibliography{main}
\end{document}